\newcommand{\kms}{\hbox{km$\,$s$^{-1}$}}
\newcommand{\cmt}{\hbox{cm$^{-3}$}}
\newcommand{\Apix}{\hbox{\AA$\,$pix$^{-1}$}}
\newcommand{\two}{\,{\sc ii}}
\newcommand{\three}{\,{\sc iii}}

\newcommand{\fsec}{\hbox{$.\!\!^{\rm s}$}}

\documentclass{emulateapj}

\usepackage{amsmath}
\usepackage{amssymb}
\usepackage{mathrsfs}
\usepackage{color}
\usepackage{graphicx}

\shorttitle{Paper I: Spatially resolved dynamics of M82}
\shortauthors{M.\ S.\ Westmoquette et al.}

\begin{document}
\defcitealias{mckeith93}{McK93}
\defcitealias{mckeith95}{McK95}
\defcitealias{shopbell98}{SBH98}
\defcitealias{castles91}{CMG91}
\defcitealias{westm07c}{W07c}
\defcitealias{westm09a}{Paper I} 
\defcitealias{westm09b}{Paper II}


\title{The optical structure of the starburst galaxy M82 -- I: Dynamics of the disk and inner-wind\altaffilmark{1}}
\author{M.\ S.\ Westmoquette\altaffilmark{2}\email{msw@star.ucl.ac.uk}}
\author{L.\ J.\ Smith\altaffilmark{2,3}}
\author{J.\ S.\ Gallagher III\altaffilmark{4}}
\author{G.\ Trancho\altaffilmark{5}}
\author{N.\ Bastian\altaffilmark{6}}
\author{I.\ S.\ Konstantopoulos\altaffilmark{2}}

\altaffiltext{1}{Based on observations with the Gemini and WIYN telescopes}
\altaffiltext{2}{Department of Physics and Astronomy, University College London, Gower Street, London, WC1E 6BT, UK}
\altaffiltext{3}{Space Telescope Science Institute and European Space Agency, 3700 San Martin Drive, Baltimore, MD 21218, USA}
\altaffiltext{4}{Department of Astronomy, University of Wisconsin-Madison, 5534 Sterling, 475 North Charter St., Madison WI 53706, USA}
\altaffiltext{5}{Gemini Observatory, Casilla 603, Colina el Pino S/N, La Serena, Chile}
\altaffiltext{6}{Institute of Astronomy, University of Cambridge, Madingley Road, Cambridge, CB3 0HA, UK}
%

\begin{abstract}
We present Gemini-North GMOS-IFU observations of the central starburst clumps and inner wind of M82, together with WIYN DensePak IFU observations of the inner $2 \times 0.9$~kpc of the disk. These cover the emission lines of H$\alpha$, [N\two], [S\two], and [S\three] at a spectral resolution of 45--80~\kms. The high signal-to-noise of the data is sufficient to accurately decompose the emission line profiles into multiple narrow components (FWHM $\sim$ 30--130~\kms) superimposed on a broad (FWHM $\sim$ 150--350~\kms) feature. This paper is the first of a series examining the optical structure of M82's disk and inner wind; here we focus on the ionized gaseous and stellar dynamics and present maps of the relevant emission line properties.

Our observations show that ionized gas in the starburst core of M82 is dynamically complex with many overlapping expanding 
structures located at different radii. Localised line splitting of up to 100~\kms\ in the narrow component is associated with expanding shells of compressed, cool, photoionized gas at the roots of the superwind outflow. We have been able to associate some of this inner-wind gas with a distinct outflow channel characterised by its dynamics and gas density patterns, and we discuss the consequences of this discovery in terms of the developing wind outflow.

The broad optical emission line component is observed to become increasingly important moving outward along the outflow channel, and in general with increasing height above/below the plane. Following our recent work on the origins of this component, we associate it with turbulent gas in wind-clump interface layers and hence sites of mass loading, meaning that the turbulent mixing of cooler gas into the outflowing hot gas must become increasingly important with height, and provides powerful direct evidence for the existence of mass-loading over a large, spatially extended area reaching far into the inner wind. We discuss the consequences and implications of this.

We confirm that the rotation axis of the ionized emission-line gas is offset from the stellar rotation axis and the photometric major axis by $\sim$12$^{\circ}$, not only within the nuclear regions but over the whole inner 2~kpc of the disk. This attests to the perturbations introduced from M82's past interactions within the M81 group. Finally, finding a turn-over in the stellar and ionized gas rotation curves on both sides of the galaxy indicates that our sight line, in places, extends at least half way through disk, and conflicts with the high levels of obscuration usually associated with the nuclear regions of M82.
\end{abstract}


\section{Introduction} \label{sect:intro}

M82 is the archetype starburst galaxy \citep{oconnell78, oconnell95} and one of the nearest analogues to the population of intensely star-forming galaxies that have been identified at high-$z$ \citep[e.g.][]{pettini01, shapley03}. The current ($\sim$10~Myr) starburst activity is centred on the nucleus with a diameter of $\sim$500~pc ($\sim$30$''$). Within this region there are a number of prominent, high surface-brightness clumps, first identified and labelled by \citet{oconnell78}. From \textit{Hubble Space Telescope} (\textit{HST}) imaging, each of these clumps is known to contain many hundreds of young massive star clusters \citep{oconnell95, melo05}, and it is the energy from these clusters that drives the famous H$\alpha$- and X-ray-bright, kpc-scale, superwind (\citealt{shopbell98}, hereafter \citetalias{shopbell98}; \citealt{stevens03a, strickland07}).

The bipolar outflow can be observed easily due to the galaxy's almost edge-on inclination \citep[$i \sim 80^{\circ}$;][]{lynds63, mckeith95}. The outflow is centred on regions A and C (\citetalias{shopbell98}; \citealt{ohyama02}), and is composed of a complex morphology of loops and filaments. Very few investigations of the wind dynamics have been carried out since line diagnostics of the emitting plasma are difficult to obtain with current detector technology. \citet[][hereafter \citetalias{mckeith95}]{mckeith95} presented deep optical and near-infrared long-slit observations of emission and absorption lines along M82's minor axis. Clear line splitting in their emission lines was interpreted as originating from the front and back walls of a cone-shaped structure. Inflection points in the position-velocity graph define the edge of a $\sim$300~pc ``energy injection region'', within which the separation of the components increases rapidly (i.e.\ the outflow is accelerating). Beyond this, the optical/near-IR line-emitting outflow appears to reach a $\sim$constant deprojected velocity of $\sim$600~\kms. \citetalias{mckeith95} inferred, through the difference in the strength of the red and blue components, that the outflow is inclined such that the southern cone is directed toward the observer \citepalias[][their fig.~5]{mckeith95}.

What is also known is that dust in the M82 halo scatters emission from a central source, producing an extended, smooth H$\alpha$ reflection nebula (\citealt{schmidt76, bland88}; \citetalias{shopbell98}). The emission from this faint exponential halo is broad ($\sim$300~\kms), has a radial velocity close to $v_{\rm sys}$, and exhibits a high [N\two]/H$\alpha$ ratio \citep{bland88}. From their H$\alpha$ linear polarisation maps, \citet{scarrott91} inferred that the central illuminating source of this scattered radiation must be $<$$4''$ ($<$70~pc) in diameter and located close to the peak of the IR emission longward of 10~$\mu$m (close to/within region C). This location, however, is not coincident with the dynamical centre or the position of the 2.2~$\mu$m peak usually associated with the galaxy nucleus. The presence of this scattered component evidently complicates any interpretation of the outflow dynamics determined from optical emission lines.

The major drawback of long-slit studies is the highly restricted nature of their spatial coverage. \citetalias{shopbell98} examined the M82 wind using spatially resolved Fabry-P\'{e}rot observations. By comparing their observations to simple geometric models, they determined that the (southern) outflow is cylindrical out to $\sim$500~pc, with a base diameter of $\sim$350~pc, then opens out to a cone of opening angle $\sim$25$^{\circ}$ (see their fig~11b). The inner cylindrical (or pipe-like) region is associated with the aforementioned energy injection zone. Although the line splitting is clear, they found that the optical emission is by no means distributed evenly over the surfaces of the cones, nor emanates smoothly from the entire starburst region. Clearly evident from their narrow-band imaging and line ratio maps \citep[see also][]{ohyama02}, the southern wind can be traced directly back to starburst clumps A and C, and instead of solid cones, the optically-emitting gas must be excited along filaments in the walls of the cone-like outflow \citep[cf.\ NGC 3079;][]{cecil01}. Evidence that the outflow rotates in the same sense as the disk up to heights of $>$1~kpc (\citetalias{shopbell98}; \citealt{greve04}) implies that rotating disk material has been entrained and diverted into the wind.

Radial velocities along the major axis have been studied many times in the optical/near-IR (e.g.\ \citealt{burbidge64, saito84, mckeith93, sofue98}; \citetalias{shopbell98}; \citealt{castles91, westm07c}; Konstantopoulos et al.\ in prep.), IR \citep[e.g.][]{telesco91, larkin94, achtermann95} and radio \citep[e.g.][]{shen95, wills00, seaquist01, walter02, r-r04}. Overall, these studies have found an inner $\sim$$5''$ region with a steep velocity gradient, an outer region $>$$\pm$$25''$ of constant (or slowly increasing) velocities, and a transition region where the radial velocities turn over. \citet[][hereafter \citetalias{mckeith93}]{mckeith93} were the first to notice that the gradient of the major axis velocity curve appears to increase with wavelength (from [O\two]~3729~\AA\ to Ne\two~12.8~$\mu$m). They interpreted this as arising from the fact that longer wavelengths probe deeper into the galaxy, where for the same projected position the \textit{radial (line-of-sight) component} of the orbital motion is greater.

The increasing wealth of velocity measurements have lead to the discovery of a stellar bar \citep{telesco91, achtermann95}. Through comparison to models, the bar was found to possess two families of intersecting orbits -- an outer set of so-called $x_{1}$-orbits (which form the main bar structure, and are traced mostly by the cooler neutral gas), and an inner set of perpendicularly oriented $x_{2}$-orbits contained within the inner $\sim$$5''$ region, and are associated with the ionized gas \citep{wills00, greve02}. The latter orbit family arises only in the presence of an inner Lindblad resonance \citep{athanassoula92a}. The interaction between the $x_{1}$ and $x_{2}$-orbit families has resulted in the build up of a torus of molecular and ionized gas and dust identified in radio and IR imaging \citep{achtermann95, weis01}. By combining \textit{HST} STIS spectroscopy and imaging, \citet[][hereafter \citetalias{westm07c}]{westm07c} suggested a possible 3D geometry for the bar with respect to the well-known features of the M82 starburst.

\citetalias{westm07c} were able to identify multiple components in the bright emission lines of H$\alpha$ and [N\two] in the central regions of M82, including a very broad (200--300~\kms) feature underlying the main narrow component. Broad emission line wings have been detected in a number of other nearby starburst galaxies (e.g.~\citealt{izotov96, homeier99, marlowe95, mendez97}; \citealt*{sidoli06}; \citealt{vanzi06}). However, due to mismatches in spectral and spatial resolution and in the specific environments observed, the nature of the energy source for these broad lines has been contested, resulting in a large number of possible explanations being proposed. Recent detailed IFU (integral field unit) studies of the ionized ISM in NGC 1569, however, have shed a considerable amount of light on this problem \citep{westm07a, westm07b, westm08}.

By mapping out the properties of the individual line components (including the broad underlying component), \citet{westm07a} identified a number of correlations that allowed them to determine the likely origin of the broad component. They concluded that the evaporation and/or ablation of material from cool interstellar gas clouds caused by the impact of the high-energy photons and fast-flowing cluster winds \citep{pittard05} produces a highly turbulent layer on the surface of the clouds \citep{slavin93, binette99} from which the emission arises. \citetalias{westm07c} showed that this explanation is also applicable to the broad lines in M82. They argued that since the high pressure ISM is highly fragmented, with many small clouds well mixed in with the star clusters, there are many cloud surfaces with which the copious ionizing photons and fast winds can interact.

Clearly, the structure and dynamics of the M82 system are complex. They vary both on the large and small scales, and are complicated by the inflows and outflows caused by the bar and starburst. Long slit observations have hitherto contributed a great deal to our knowledge of the rotation and outflow dynamics, but are of limited use in the long-term due to their restrictive linear spatial coverage. We have, therefore, obtained a set of high resolution, integral field, optical and near-IR spectra of the nuclear regions and disk of M82 at two complimentary spatial scales. With the Gemini GMOS-IFU we have observed the brightest starburst clumps and part of the inner wind region in detail with five $7\times 5$~arcsec pointings. With the WIYN DensePak IFU we observed four partially overlapping regions covering the central $\sim$2~kpc of the disk. These two datasets together provide a wealth of information on the state of the ionized gas from both nebular and stellar origins, and present a unique opportunity to study the nuclear starburst and its wider environment. We have also obtained a set of very deep narrow-band images of the superwind with the WIYN telescope in order to investigate the large-scale structure of wind outflow.
 
In this paper we present the IFU datasets and describe in detail the reduction and analysis techniques employed. We then focus on the gaseous and stellar kinematics (line widths and radial velocities), and discuss and contrast how the dynamics of the nuclear regions connect with those of the large-scale disk and the starburst system as a whole. In a companion paper (Westmoquette et al.\ in prep.; hereafter Paper~II), we will present the second half of our analysis of the IFU data, focussing on the properties of the ionized gas in the wind, including a discussion of the gas densities and excitations. In a forthcoming contribution (Gallagher et al.\ 2008; hereafter Paper~III) we will present new deep narrow-band imaging of the wind and put forward an morphological picture of the whole wind system.

In this work, we adopt a distance to M82 of 3.6~Mpc \citep{freedman94}, meaning $1'' = 17.5$~pc, a systemic velocity, $v_{\rm sys}$, of $+200$~\kms\ \citepalias{mckeith93}, and a photometric major axis position angle (PA) of $65^{\circ}$ (\citealt{vaucouleurs91}; \citetalias{shopbell98}).


\section{Observations and data reduction} \label{sect:obs}

\subsection{Gemini observations} \label{sect:GMOS_obs}
In February 2006 and February 2007 queue-mode observations using the Gemini-North Multi-Object Spectrograph (GMOS) Integral Field Unit \citep[IFU;][]{allington02} were obtained covering six regions near the centre of M82 (programme IDs: GN-2006A-Q-38, and GN-2007A-Q-21, PI: L.J.\ Smith), with 0.3--0.8~arcsec seeing (5--14~pc at the distance of M82). A nearby bright star was used to provide guiding and tip-tilt corrections using the GMOS on-instrument wave front sensor \mbox{(OIWFS)}.

Opting for two-slit mode gave us an IFU field-of-view (FoV) of $7\times 5$~arcsecs ($\sim$$50\times 35$~pc at the distance of M82) sampled by 1000 contiguous hexagonal fibres of $0\farcs2$ diameter. An additional block of 250 fibres (covering $2.5\times 1.7$~arcsecs), offset by $1$ arcmin from the object field, provided a dedicated sky view. We took two dithered exposures at each position (dither offset = 0.5 arcsec parallel to short axis of IFU), with integration times between 1200 and 2400 secs each, and used the R831 grating to give a spectral coverage of 6100--6790~\AA. This allowed us to cover the nebular diagnostic lines of H$\alpha$, [N\two]$\lambda\lambda 6548,6583$ and [S\two]$\lambda\lambda 6716,6731$ at a dispersion of 0.34~\Apix. Table~\ref{tbl:gmos_obs} lists the coordinates, position angles (PAs) and exposure times for each position.

The GMOS spectrograph is fed by optical fibres from the IFU which reformats the arrangement of the spectra for imaging by the detector. This is composed of three 2048\,$\times$\,4068 CCD chips separated by small gaps. In order to remove the pixel-to-pixel sensitivity differences, and enable the wavelength and flux calibration of the data, a number of bias frames, flat-fields, twilight flats, arc calibration frames, and observations of the photometric standard stars G191-B2B and Hiltner 600, were taken contemporaneously with the science fields.

\begin{figure}
\centering
\plotone{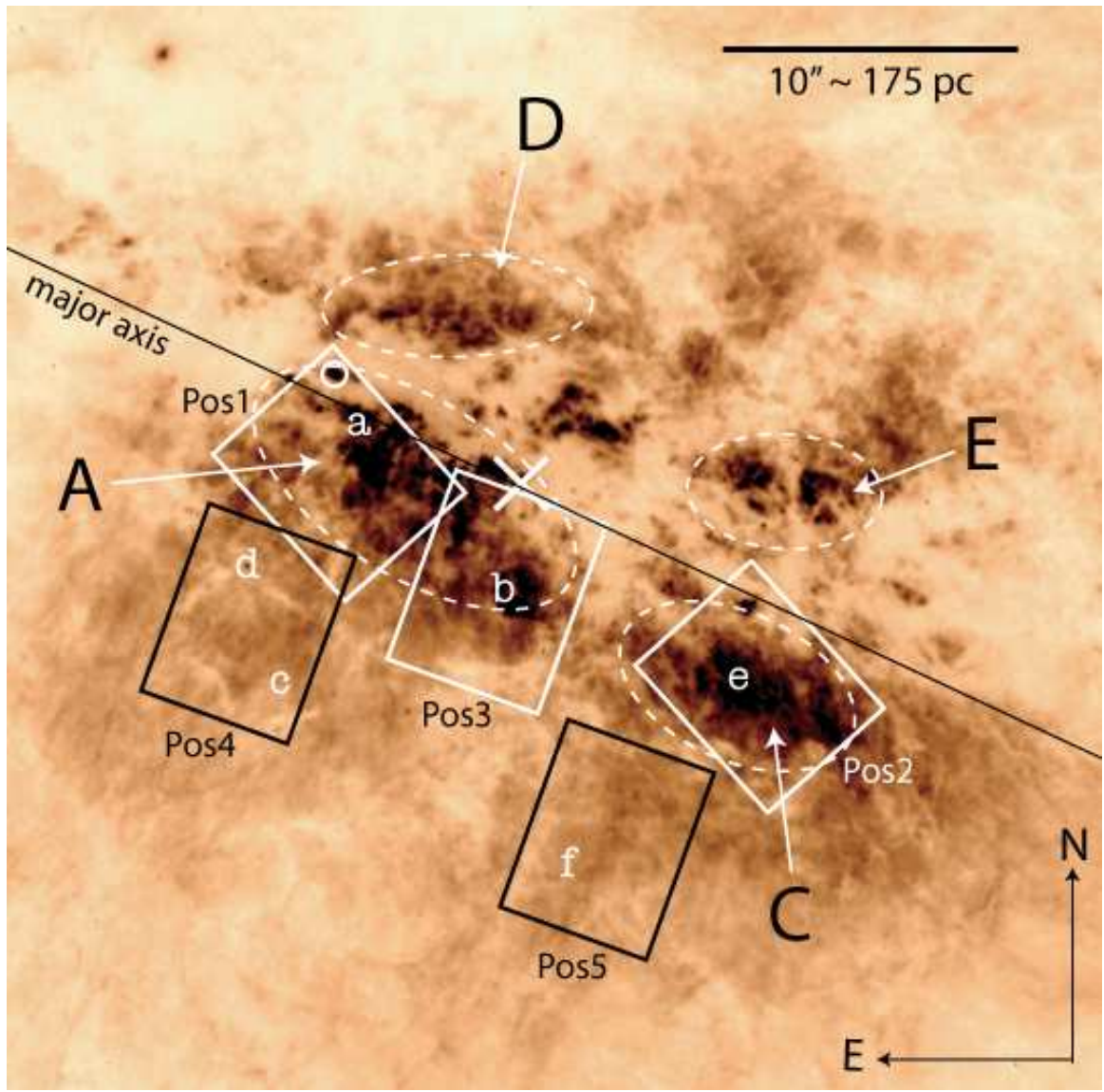} 
\caption{\textit{HST}/ACS F656N image of the central regions of M82, showing the position of our five 7$\times$5~arcsec IFU fields. The brightest starburst clumps are labelled and outlined with dashed lines. The 2.2~$\mu$m nucleus \citep{lester90} is marked by a white cross, the major axis by a solid line, and the position of the cluster M82-A1 with a white circle. The white lowercase letters indicate the location from which the H$\alpha$ line profiles shown in Fig.~\ref{fig:egfits} were extracted.}
\label{fig:GMOSfinder}
\end{figure}

The IFU positions were chosen to cover selected areas of the disturbed ionized interstellar medium in the inner region of M82, near the roots of the wind outflow; positions 1 and 2 cover the bright starburst clumps A and C \citep[nomenclature from][]{oconnell78}, whereas positions 3, 4 and 5 were aligned perpendicularly to the major axis and cover the inter-clump region and the outflowing gas to the south. In Fig.~\ref{fig:GMOSfinder}, we show the position of the IFU fields on an archive \textit{HST}/ACS F656N image (GO 9778, P.I.\ L.\ Ho). Super star cluster (SSC) A1 \citep{smith06} is located on the north-easternmost corner of position 1, as indicated with a white circle. The sixth IFU field was positioned over the star cluster M82-F located $\sim$500~pc to the west of the nucleus, and, since an analysis of these data is presented in \citet{bastian07}, will not be discussed here. 

\begin{table}
\centering
\caption {Gemini GMOS-IFU observations.}
\label{tbl:gmos_obs}
\begin{tabular}{c c r @{\hspace{0.2cm}} l c r @{ $\times$ } l }
\hline
Pos. & Date & \multicolumn{2}{c}{Coordinates} & PA & \multicolumn{2}{c}{Exp.\ Time} \\
No. & & \multicolumn{2}{c}{(J2000)} & ($^{\circ}$) & \multicolumn{2}{c}{(s)} \\
\hline 
1 & 06/2/06 & $09^{\rm h}\,55^{\rm m}\,53\fsec29$ & $69^{\circ}\,40'\,46\farcs8$ & 312 & 1200 & 2 \\
2 & 14/2/07 & $09^{\rm h}\,55^{\rm m}\,50\fsec69$ & $69^{\circ}\,40'\,39\farcs6$ & 312 & 1200 & 2 \\
3 & 14/2/07 & $09^{\rm h}\,55^{\rm m}\,52\fsec31$ & $69^{\circ}\,40'\,42\farcs9$ & 250 & 2100 & 2 \\
4 & 14/2/07 & $09^{\rm h}\,55^{\rm m}\,53\fsec77$ & $69^{\circ}\,40'\,41\farcs6$ & 250 & 2400 & 2 \\
5 & 15/2/07 & $09^{\rm h}\,55^{\rm m}\,51\fsec64$ & $69^{\circ}\,40'\,34\farcs3$ & 250 & 2400 & 2 \\
\hline\\
\end{tabular}
\end{table}

\subsubsection{Reduction} \label{sect:GMOSreduction}
The field-to-slit mapping for the GMOS IFU reformats the layout of the fibres of the sky to one long row containing blocks of 100--150 object fibres interspersed by blocks of 50 sky fibres. In this way, all 1250 spectra can be recorded on the detector simultaneously.

Basic data reduction was performed following the standard Gemini reduction pipeline (implemented in \textsc{iraf}\footnote{The Image Reduction and Analysis Facility ({\sc iraf}) is distributed by the National Optical Astronomy Observatories which is operated by the Association of Universities for Research in Astronomy, Inc. under cooperative agreement with the National Science Foundation.}). Briefly, a trace of the position of each spectrum on the CCD was first produced from the flat field. Then throughput correction functions and wavelength calibration solutions were created and applied to the science data, before the individual spectra were extracted using the flat-field trace to produce a data file containing 1250 reduced spectra, each one pixel in width, and ordered by the position they were recorded on the CCD (hereafter the `fibre order'). After extraction of the spectra, the $x$ and $y$ spatial units are termed `spaxels' to differentiate from `pixel', which refers to the CCD. The final steps of the pipeline were to clean the cosmic-rays using \textsc{lacosmic} \citep{vandokkum01}, subtract an averaged sky spectrum (computed from the separate sky field), and apply a flux calibration derived from observations of the standard star (G191-B2B for the Feb 06 run, and Hiltner600 for the Feb 07 run). Note that although the one arcmin separation of the GMOS sky and object fields (see above) was not enough to place the sky field on a completely dark region of the sky, the contamination from emission in the halo of M82 was found to be $<$0.5\%. Separation of the sky spectra from the science data resulted in a data file formed of 1000 fully reduced spectra.

The data were converted into standard `cube' format using \textsc{gfcube} with no interpolation. To combine the individual dithered exposures for each position, we used a custom \textsc{pyraf} script. After the required interpolation was applied to combine the fields, the result was a data file containing 825 spaxels on which we could begin our analysis.

In order to determine an accurate measurement of the instrumental contribution to the line broadening, we fitted single Gaussians to three isolated arc lines on a wavelength calibrated arc exposure associated with position 1, for all 1495 apertures (science+sky). The average instrumental broadening (velocity resolution; FWHM) of the final dataset is $1.26\pm 0.06$~\AA, or $59\pm 3$~\kms\ \citep[this is consistent with our previous work with this instrument;][]{westm07a}. The wavelength range covered is small enough such that instrumental resolution does not vary significantly from the blue to the red extremes. Similarly, differential atmospheric refraction does not affect our data in any significant way.

\subsection{WIYN observations} \label{sect:DP_obs}
DensePak \citep{barden98} is a small fibre-fed integral field array attached to the Nasmyth focus of the WIYN (Wisconsin, Indiana, Yale and NOAO) 3.5-m telescope. It has 91 fibres, each with a diameter of 300~$\mu$m or $3''$ on the sky; the fibre-to-fibre spacing is 400~$\mu$m making the overall dimensions of the array $30\times 45$~arcsecs. Four additional fibres are offset by $\sim$$60''$ from the array centre and serve as dedicated sky fibres. The format of the DensePak fibre array on the sky, including the sky fibres is shown in \citet[][]{sawyer97}. At the time of observation, there were 5 damaged and unusable fibres in the main array (36, 40, 46, 59 and 93) making a usable total of 86. DensePak's fibre bundle is reformatted into a `pseudo-slit' to feed the WIYN bench-mounted echelle spectrograph. This spectrograph is part of the Hydra multi-object fibre positioner instrument, and uses a T2KC $2048\times 2048$ CCD detector. 

On 14th April 2001, four fields covering the stellar disk of M82 were observed with DensePak. Unfortunately, due to a telescope malfunction, the precise coordinates for the fourth position were not recorded at the time of observation. To recover approximate coordinates we have matched reconstructed images in the continuum and H$\alpha$ bands to WIYN and \textit{HST} imaging. The resulting certainty in the coordinates for position 4 is degraded accordingly. The 860~line~mm$^{-1}$ grating gave a wavelength range of 5820--6755~\AA\ with a dispersion of 0.46~\Apix, allowing us access to a number of optical nebular lines (hereafter referred to as the H$\alpha$ grating). The first three positions were also observed with a second grating setting with a wavelength range of 7745--9700~\AA\ and dispersion of 0.96~\Apix, providing access to the near-IR Ca\two\ stellar absorption lines ($\lambda\lambda\lambda$8498,8542,8662) and a number of additional nebular emission lines (hereafter referred to as the Ca\two\ grating). Details of the observations are given in Table~\ref{tbl:dp_obs}. A number of bias frames, flat-fields and arc calibration exposures were also taken together with the science frames.

\begin{table}
\caption{WIYN DensePak observations. Coordinates refer to the array centre (fibre 42).}
\label{tbl:dp_obs}
\centering
\centering
\begin{tabular}{lllc}
\hline
Position & \multicolumn{2}{c}{(J2000)} & Exposure Time \\
& \multicolumn{1}{c}{RA} & \multicolumn{1}{c}{Dec} & (s) \\
\hline
H$\alpha$ \textit{grating} \\
1 & $9^{\rm h}\,55^{\rm m}\,53\fsec53$ & $69^{\circ}\,40'\,43\farcs14$ & $3\times 1200$ \\
2 & $9^{\rm h}\,55^{\rm m}\,48\fsec56$ & $69^{\circ}\,40'\,32\farcs20$ & $3\times 1200$ \\
3 & $9^{\rm h}\,56^{\rm m}\,04\fsec11$ & $69^{\circ}\,41'\,11\farcs57$ & $3\times 1200$ \\
4 & $9^{\rm h}\,55^{\rm m}\,55\fsec7$ & $69^{\circ}\,40'\,57''$ & $1\times 1200$ \\
Ca\two\ \textit{grating} \\
1 & $9^{\rm h}\,55^{\rm m}\,53\fsec53$ & $69^{\circ}\,40'\,43\farcs14$ & $2\times 1200$ \\
2 & $9^{\rm h}\,55^{\rm m}\,48\fsec56$ & $69^{\circ}\,40'\,40\farcs75$ & $1\times 1200$ \\
3 & $9^{\rm h}\,56^{\rm m}\,04\fsec11$ & $69^{\circ}\,41'\,11\farcs57$ & $2\times 1200$ \\
\hline\\
\end{tabular}
\end{table}

\subsubsection{Reduction} \label{sect:DPreduction}
Basic reduction was done using the {\sc ccdproc} task within the {\sc noao} {\sc iraf} package. 
Instrument-specific reduction was then achieved using the {\sc hydra} tasks, also within the {\sc noao} package. The first step was to run {\sc apfind} on the master flat-field exposure to automatically detect the individual spectra on the CCD frame. The output of this task is an aperture identification table containing the trace of the dispersed light from each fibre, which can be applied to the science frames to extract out the individual spectra. The task {\sc dohydra} was then run to perform the flat-fielding and wavelength calibration. The datafile at this stage contained 90 reduced and wavelength calibrated spectra (one for each fibre in use, including the four sky fibres).

Even with the sky fibres separated from the main array by $60''$, extended emission from the galaxy contaminated, to varying degrees, each sky fibre in each of the positions. After some testing, we found that sky fibre 90 of position 3 was the least contaminated. We therefore formed a representative sky spectrum by averaging the data from this fibre from all three exposures, then manually removing the remaining low-level contamination (using the `deblend' function in \textsc{splot}). This sky spectrum was then subtracted from the data (after scaling to the continuum level of fibre 90 for the other positions). Cosmic-rays were cleaned from the data using {\sc lacosmic} \citep{vandokkum01}, before final combination of the individual frames using {\sc imcombine}. The final datafiles now contained 86 (sky fibres removed) reduced and sky-subtracted spectra.

In order to determine an accurate measurement of the instrumental contribution to the spectrum broadening for the two grating settings, we selected high S/N spectral lines from wavelength calibrated arc exposures that were close to the H$\alpha$ (H$\alpha$ grating) and [S\three]$\lambda$9069 (Ca\two\ grating) line in wavelength, and sufficiently isolated to avoid blends. We then fitted a Gaussian to these lines for all 86 apertures and took the average. The instrumental broadening is 44.7\,$\pm$\,3.1~\kms\ in the H$\alpha$ grating and 81.8\,$\pm$\,4.1~\kms in the Ca\two\ grating.

\begin{figure*}
\centering
\plotone{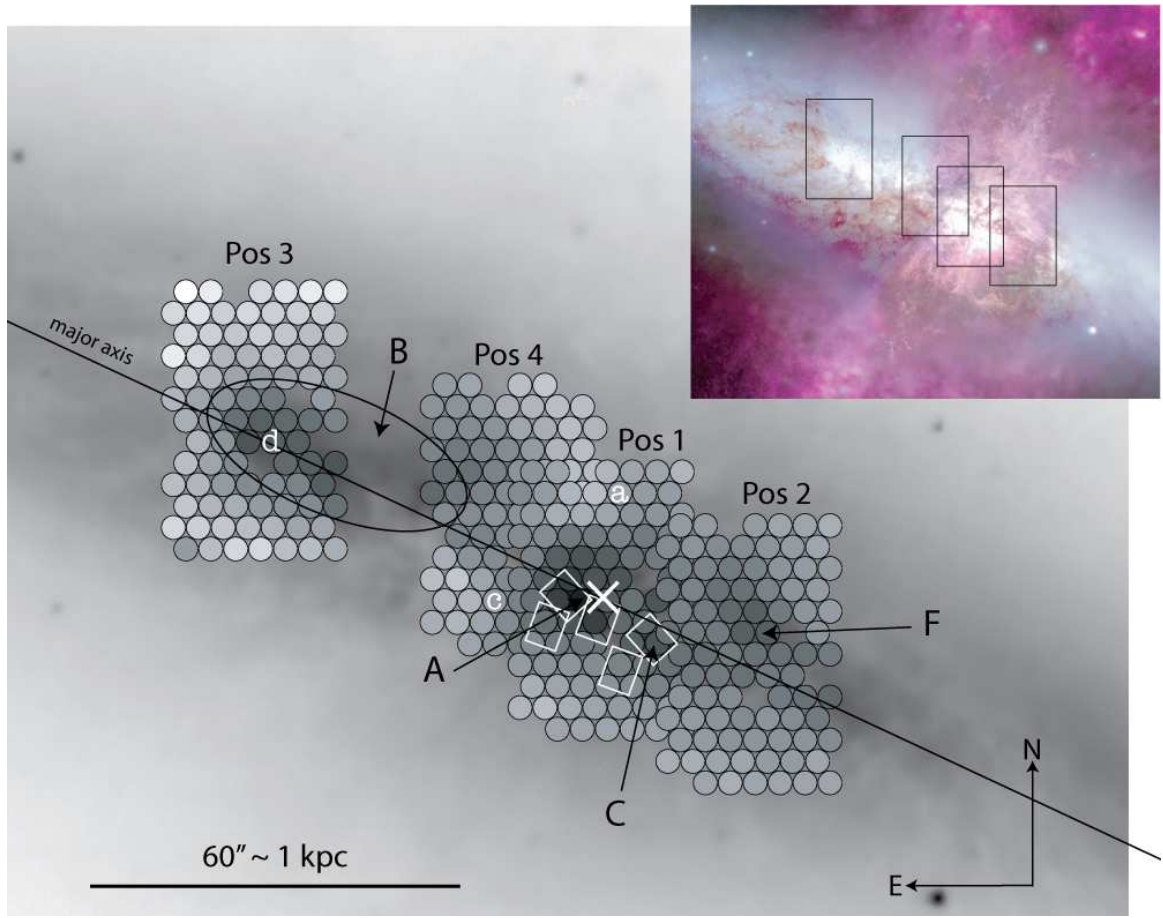} 
\caption{WIYN $R$-band image of M82 (inverted greyscale) overlaid with DensePak continuum flux maps (also inverted) for all three pointings. Some of the bright starburst clumps/clusters are marked, and the position of the 2.2~$\mu$m nucleus \citep{lester90} is shown with a white cross. The boxes show the position of the GMOS IFUs, and the galaxy major axis (PA = 65$^{\circ}$) is indicated by a solid line. The letters indicate the spaxels from which the H$\alpha$+[N\two] line profiles shown in Fig.~\ref{fig:dp_fits} were extracted (letter `b' is located at the end of the region A arrow, but has been omitted for clarity).
The inset shows an \textit{HST}+WIYN composite (Paper~III) covering the same area as that shown in the main image, and includes narrow-band emission from the wind outflow. The outlines of the DensePak fields are shown for comparison. \textit{[A colour version of this figure is included in the on-line version.]} }
\label{fig:DPfinder}
\end{figure*}

\begin{figure}
\centering
\plotone{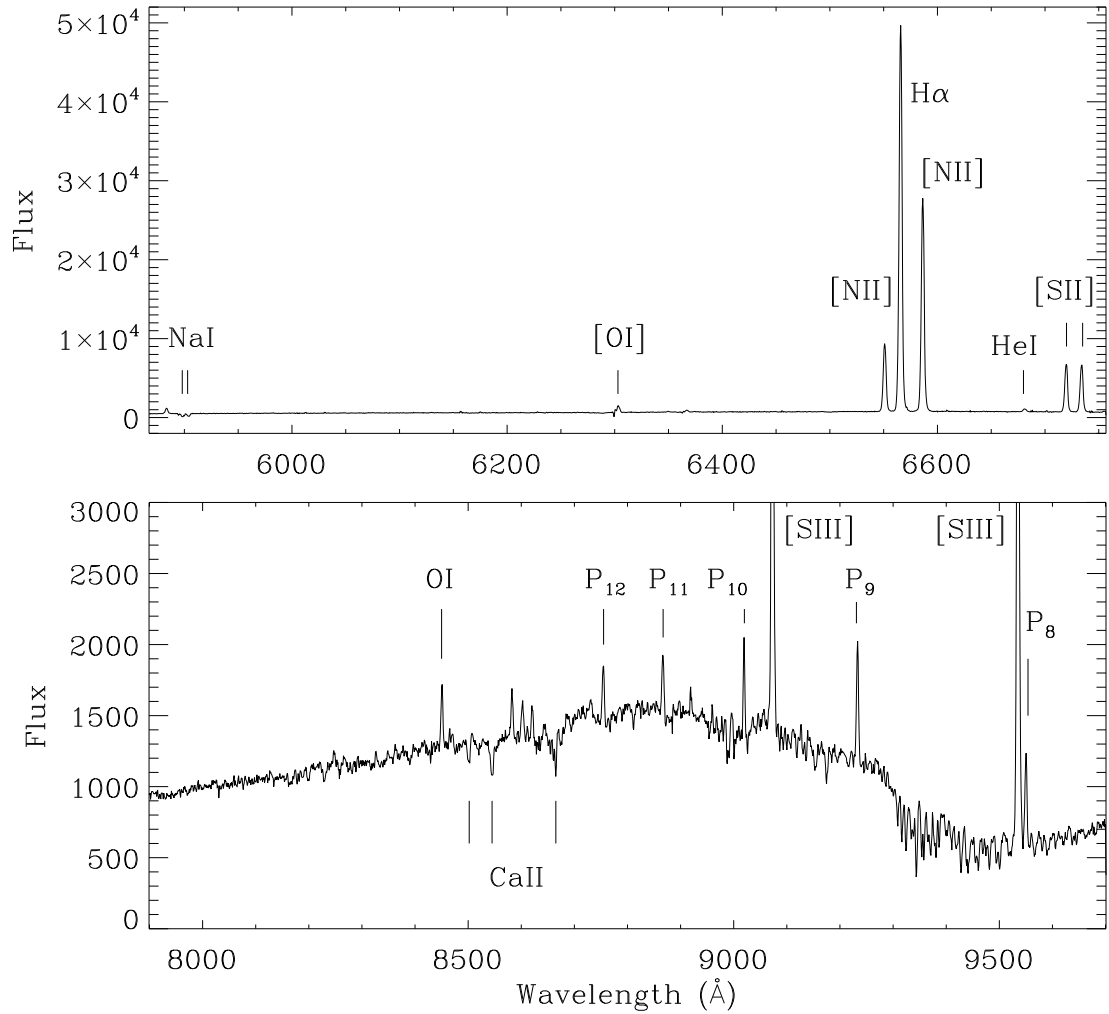} 
\caption{Example DensePak spectra from fibre 35 in position 1 showing the full wavelength ranges observed using both gratings. Emission and absorption lines are labelled; the continuum shape in the red grating arises because the spectra are not flux calibrated.}
\label{fig:spec_eg_DP}
\end{figure}

Fig.~\ref{fig:DPfinder} shows a reconstructed image of the three combined DensePak fields in the line-free continuum region 6645--6660~\AA, overlaid on a WIYN $R$-band image of the nuclear regions. The inset shows the position of the DensePak fields on an \textit{HST}+WIYN narrow-band composite (Paper~III), for comparison to the H$\alpha$ emission from the wind. Position 1 was centred on region A of the starburst, position 2 samples the region to the west of the main starburst nucleus, including the area surrounding cluster F, and position 3 was centred on region B. Position 4 samples the eastern part of the nuclear starburst region, including part of the north-south dust lane. In many of the fibres near the starburst nucleus, the signal-to-noise (S/N) is very high ($>$500 at the centre of position 1). Two example spectra from each of the grating settings are shown in Fig.~\ref{fig:spec_eg_DP}.

\subsection{Decomposing the line profiles} \label{sect:line_profiles}

Following the methodology employed by \citet{westm07a,westm07b}, we fitted multiple Gaussian profile models to each emission line using an \textsc{idl} $\chi^{2}$ fitting package called \textsc{pan} \citep[Peak ANalysis;][]{dimeo}, to quantify the gas properties observed in each IFU field. A detailed description of the program and our fitting methods can be found in \citet{westm07a}.

Emission lines are detected in every fibre of both the GMOS and DensePak datasets. The high S/N and spectral resolution of our data have allowed us to quantify the line profile shapes of these lines to a high degree of accuracy. In general, we find the emission lines to be composed of a bright, narrow component (hereafter C1; this was sometimes split into two narrow components, hereafter C1 and C3) overlaid on a fainter, broad component (hereafter C2). Each line in each of the 825 spaxels (GMOS IFUs) and 86 spaxels (DensePak IFUs) was therefore fitted using a single, double, and triple Gaussian component initial guess. Line fluxes were constrained to be positive and widths to be greater than the instrumental contribution to guard against spurious results. In some regions of our DensePak data, we can also identify an H$\alpha$ absorption component of stellar origin (hereafter C4). Here we add an additional Gaussian absorption component, with a flux constrained to be negative. To fit the H$\alpha$ and [N\two]$\lambda 6583$ lines, for each component we constrained the wavelength difference between the two Gaussian models to be equal to the laboratory difference, and FWHMs to equal one another. This was also the approach taken for fitting the [S\two] doublet. Multi-component fits were run several times with different initial guess configurations (widths and wavelengths) in order to account for the varied profile shapes. However, we note that the $\chi^2$ minimisation routine employed by PAN is very robust with respect to the configuration of the fit initial guess. 

In order to determine which fits were correct, and how many Gaussian components best fit an observed profile, we applied a number of tests and filters. Firstly, to select which result to keep of the multi-component fits made with varied initial guess parameters, we simply chose the fit with the lowest $\chi^{2}$ value. Secondly, to determine how many Gaussian components best fit an observed profile (one, two or three), we used the statistical F-test. The F-distribution function allows one to calculate the significance of a variance ($\chi^{2}$) increase that is associated with a given confidence level, for a given number of degrees of freedom. The test will output the minimal increase of the $\chi^{2}$ ratio that would be required at the given confidence limit for deciding that the two fits are different. If the $\chi^{2}$ ratio is higher than this critical value, the fits are considered statistically distinguishable, and the one with a lower $\chi^{2}$ is selected. However, this test only tells us which of the fits (single, double or triple component) is most appropriate for the corresponding line profile. Experience has taught us that a number of physically motivated tests are needed to, firstly filter out well-fit but physically improbable results, and, secondly assess which Gaussian profile belongs to which physical line component.

For a fit to be accepted, we set the criteria that the measured FWHM had to be greater than the associated error on the FWHM result (a common symptom of a bad fit). Of a triple-Gaussian fit, we specified that the broadest component should be assigned to component 2 (C2), and after that, the brightest to be component 1 (C1) and the faintest narrow component to component 3 (C3). 
With double-Gaussian fits, we assigned C2 to be the broader of the two components and C1 to the other, regardless of their velocity difference. This consistent approach helped limit the confusion that might arise during analysis of the results where discontinuous spatial regions might arise from incorrect component assignments.

Experience has shown us that the errors that \textsc{pan} quotes on its fit results are an underestimate of the true uncertainties. This is certainly true once uncertainties on our post-fitting tests and filters are taken into account. We can therefore estimate more realistic errors through a visual re-inspection of the profile+fit after knowing which one was selected by our tests, and by taking into account the S/N of the spectrum. For the fluxes, we estimate that there is a 0.5--10 per cent error in C1 (with the range resulting mostly from the variation in S/N levels), 8--15 per cent on C2 and 10--80 per cent on the C3 flux. The presence of a third fit component may increase the errors in the C1 and C2 fluxes by 5--10 per cent, particularly where C3 is faint. We estimate that errors in FWHM range between 0.25--2~\kms\ for C1, 4--20~\kms\ for C2 and 15--25~\kms\ for C3, whereas those for radial velocities range between 0.1--3~\kms\ for C1, 2--5~\kms\ for C2 and 10--30~\kms\ for C3. Similar percentage increases in errors for C1 and C2 as for the fluxes may occur with the addition of the third component. 

The red grating of the DensePak data covered the Ca\two\ triplet stellar absorption lines. For the fitting of these lines, we adopted a different method. In order to simultaneously fit all three lines together we employed the penalised pixel fitting (pPXF) method, described in detail in \citet{cappellari04}. In brief, the method simply cross-correlates selected regions of a spectrum with a template, resulting in a measurement of the best-fit radial velocity. In our case we used the NIR Ca\two\ triplet templates from \citet{cenarro01} made available through the pPXF website\footnote{http://www.strw.leidenuniv.nl/$\sim$mcappell/idl/\#ppxf}.

\begin{figure*}
\centering
\begin{minipage}{5cm}
\includegraphics[width=5cm]{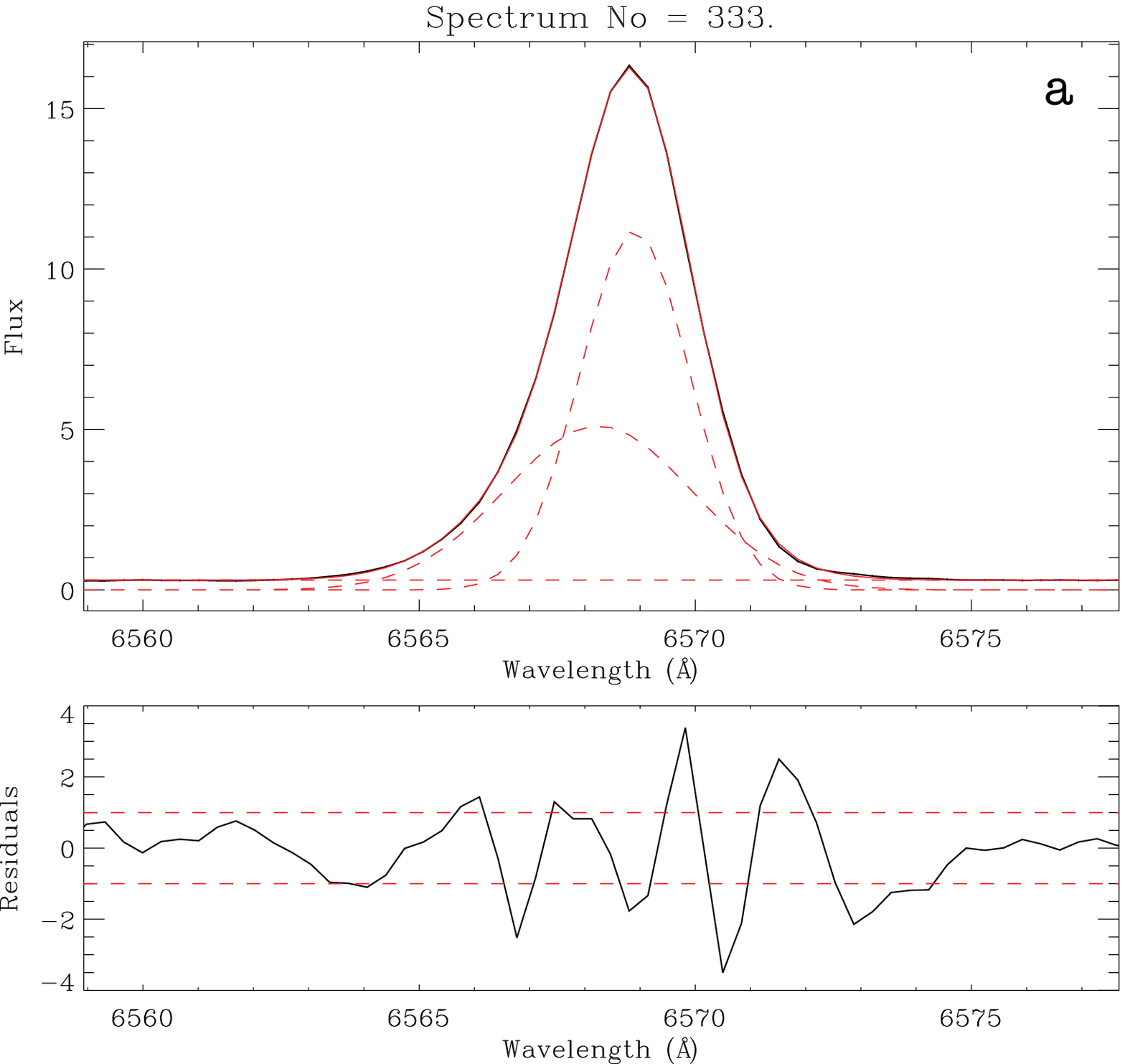} 
\end{minipage}
\hspace{0.2cm}
\begin{minipage}{5cm}
\includegraphics[width=5cm]{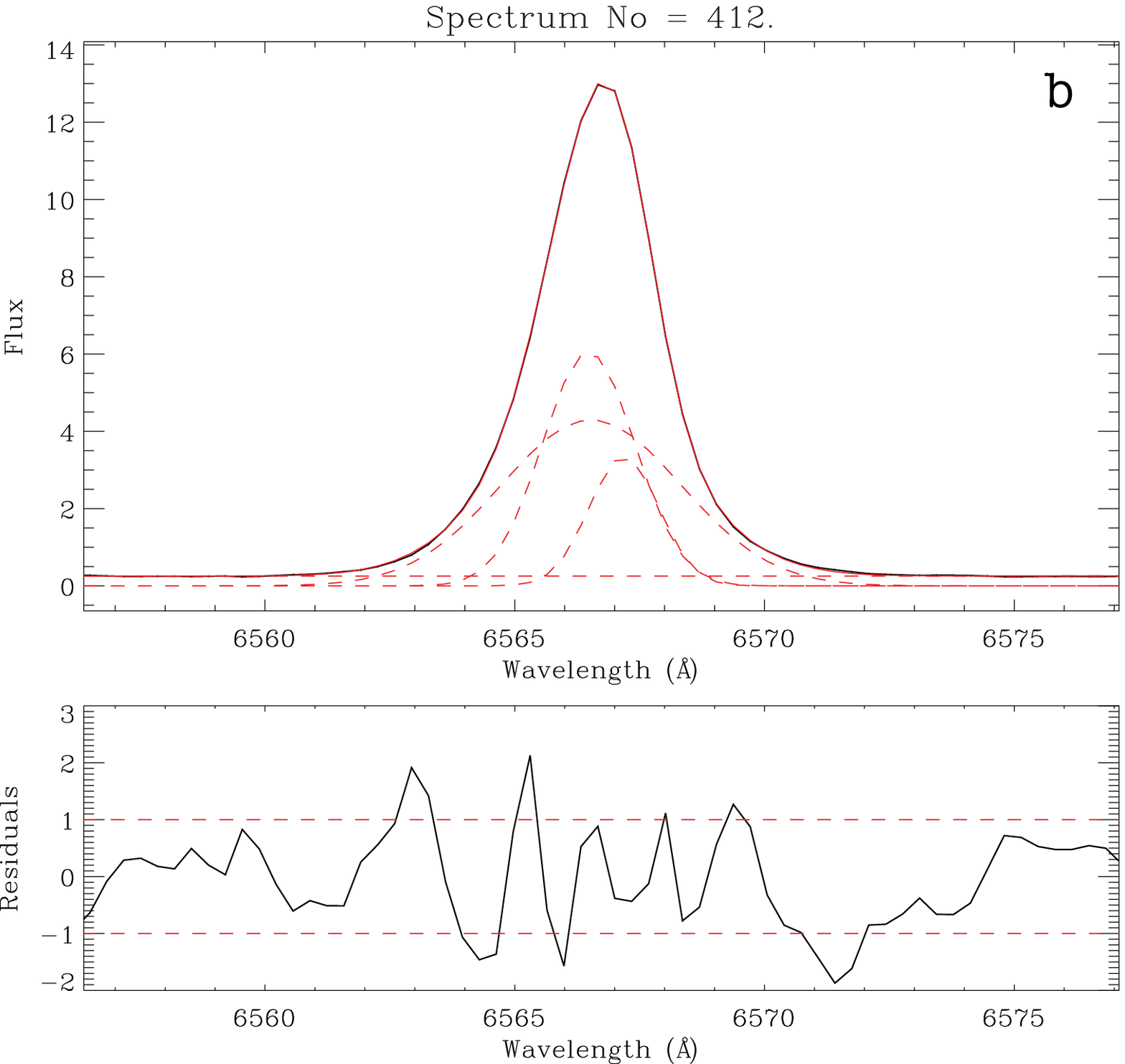} 
\end{minipage}
\hspace{0.2cm}
\begin{minipage}{5.1cm}
\includegraphics[width=5.1cm]{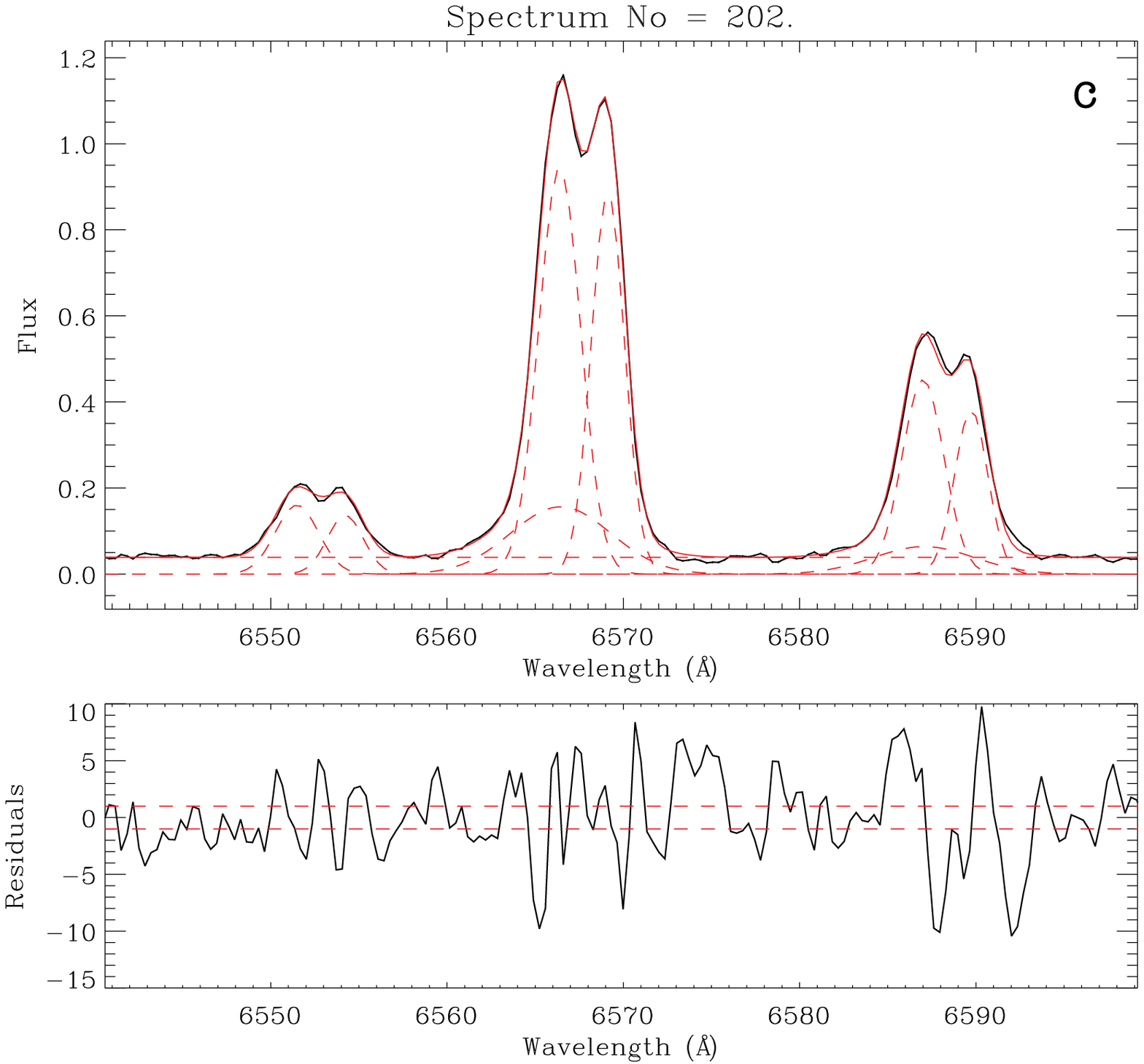} 
\end{minipage}
\begin{minipage}{5cm}
\vspace{0.3cm}
\includegraphics[width=5cm]{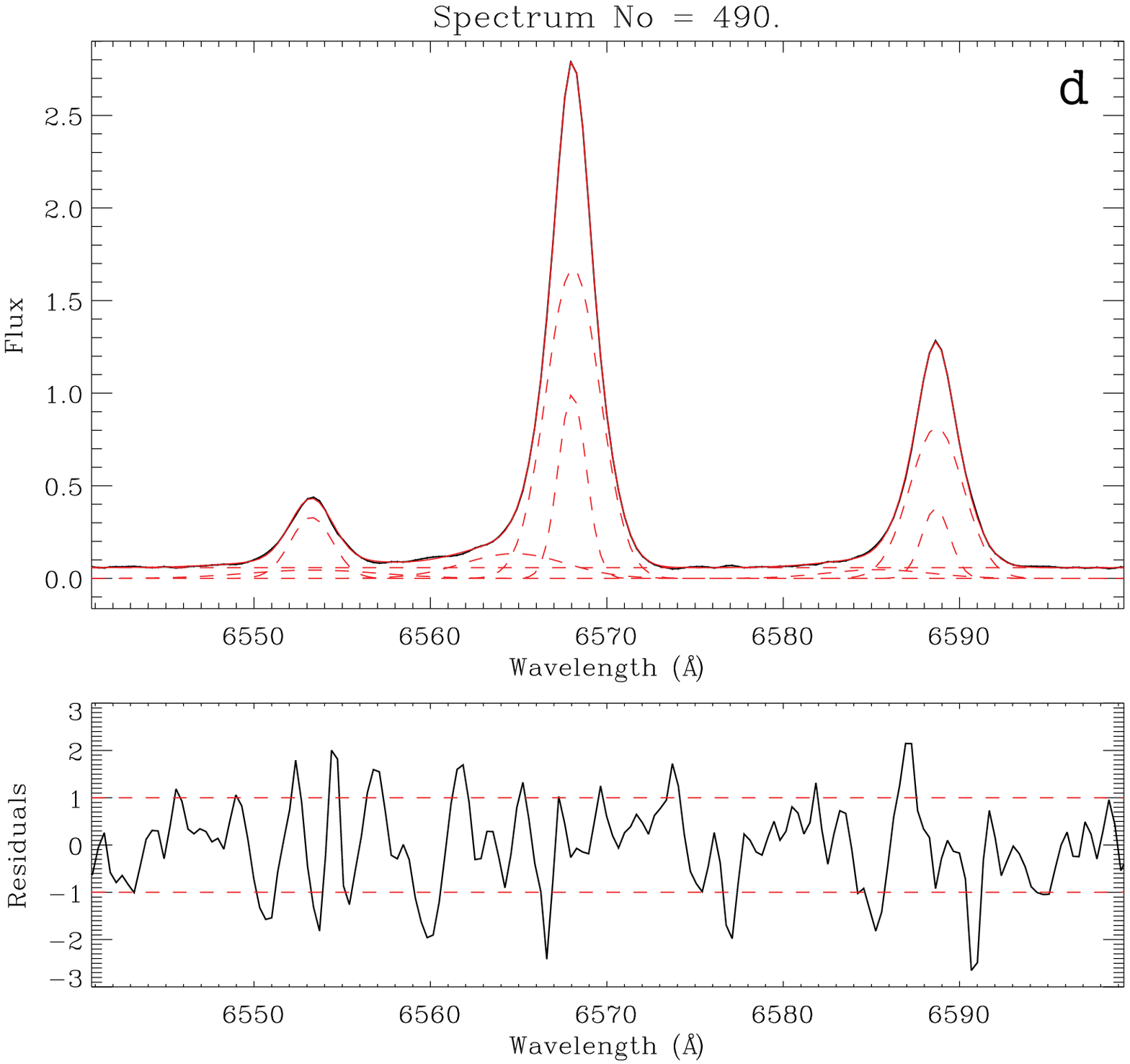} 
\end{minipage}
\hspace{0.2cm}
\begin{minipage}{5cm}
\vspace{0.3cm}
\includegraphics[width=5cm]{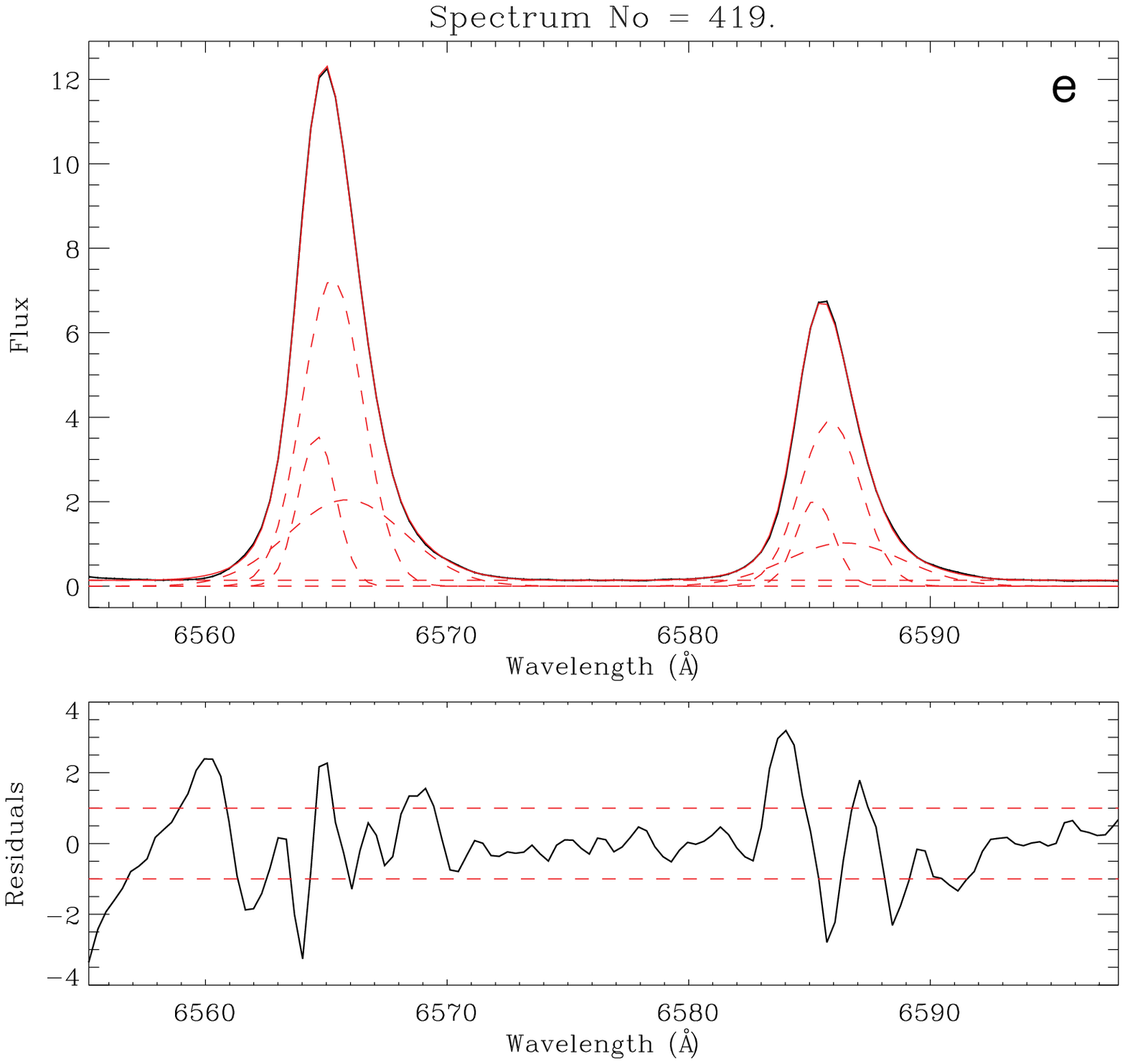} 
\end{minipage}
\hspace{0.3cm}
\begin{minipage}{5cm}
\vspace{0.3cm}
\includegraphics[width=5cm]{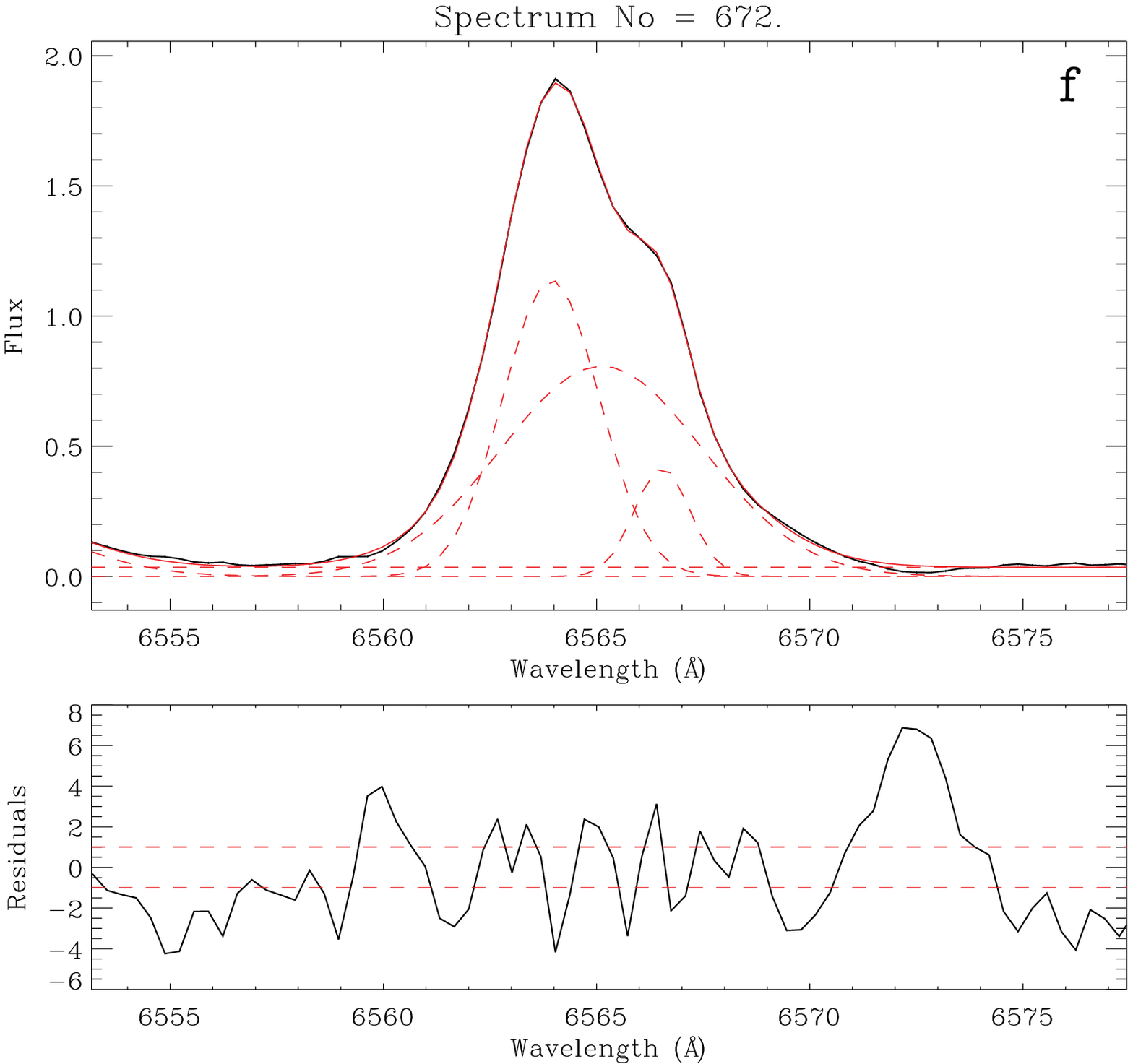} 
\end{minipage}
\caption{Example GMOS line profiles and Gaussian fits: (a) typical of the bright region A (note no line splitting in the narrow component); (b) typical of position 3 (close line splitting with a clear broad component); (c) from the south-western corner of position 4 showing clear line splitting (of 125~\kms) and the presence of a broad underlying component; (d) from the spot in the north of position 4 exhibiting very broad and blueshifted C2; (e) typical of the brightest region in the centre of region C (illustrating the consistency with which we can fit both the H$\alpha$ and [N\two] profiles); (f) from the south-east of position 5, showing a narrow component with line splitting of $\sim$50~\kms\ and a prominent broad underlying component). The units are in an arbitrary but relative intensity scale.  The fit residuals are plotted under each panel (in units of $\sigma$) with dashed guidelines at $\pm$1 (see text for further explanation). The locations from which these profiles were extracted are labelled with the corresponding letters on Fig.~\ref{fig:GMOSfinder}.}
\label{fig:egfits}
\end{figure*}

\begin{figure*}
\centering
\plotone{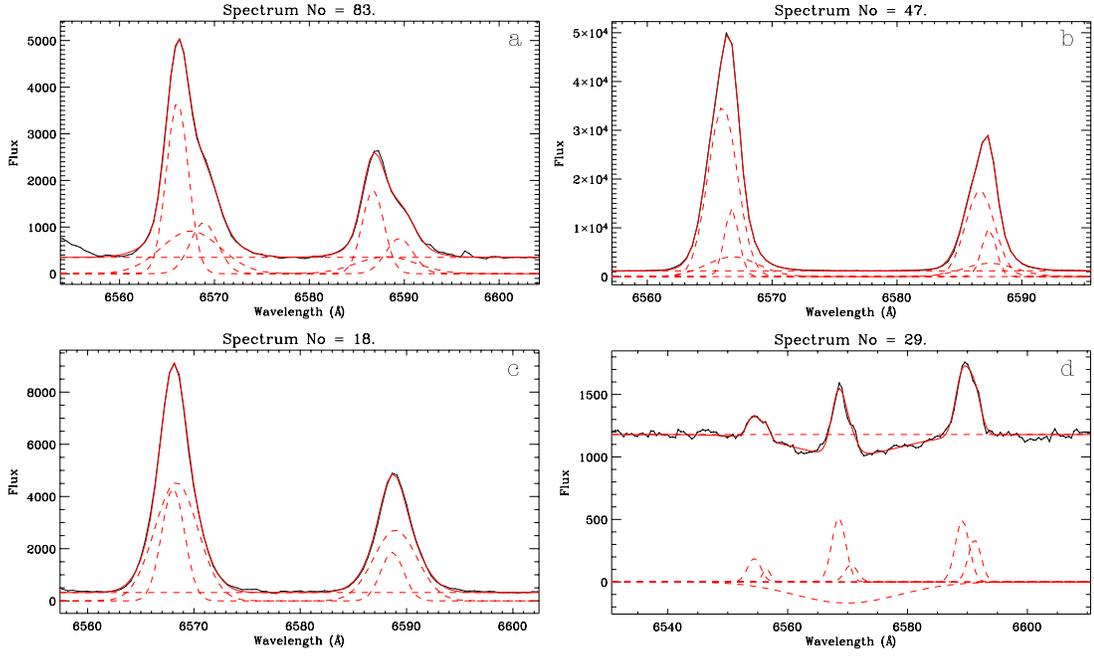} 
\caption{Example DensePak H$\alpha$ and [N\two]$\lambda 6583$ line profiles from (a) the inner-wind in the north of position 1, (b) the centre of region A, (c) to the east of region A, and (d) the centre of region B (position 3). This sample was chosen to represent a range in line profile shapes including broad, underlying emission features, second narrow peaks, and absorption profiles. The observed data are plotted as a solid black line, and the summed model fit is plotted as a solid red line. The flat continuum and individual Gaussian profiles are plotted as red dashed lines relative to the $y$-axis zero level. The units are in an arbitrary but relative intensity scale. The locations from which these profiles were extracted are labelled with the corresponding letters on Fig.~\ref{fig:DPfinder}.}
\label{fig:dp_fits}
\end{figure*}

Figs~\ref{fig:egfits} and \ref{fig:dp_fits} show a number of example H$\alpha$ (+[N\two]) line profiles and best-fitting Gaussian models from a number of regions within our GMOS and DensePak coverage (labelled with the corresponding letters in Figs~\ref{fig:GMOSfinder} and \ref{fig:DPfinder}, respectively). This sample represents the variety of different profile shapes we find across the GMOS and DensePak IFU fields, and demonstrates the high quality of the spectra and the accuracy of the line-fitting.

\section{Dynamics of the nuclear regions and inner wind} \label{sect:GMOS}

In this section we will present and describe the 2D FWHM and radial velocity maps derived from the GMOS observations. These cover five regions near the M82 nucleus, and parts of the inner-wind as shown in Fig.~\ref{fig:GMOSfinder}.

\subsection{FWHMs} \label{sect:GMOS_FWHM}

\begin{figure*}
\centering
\plotone{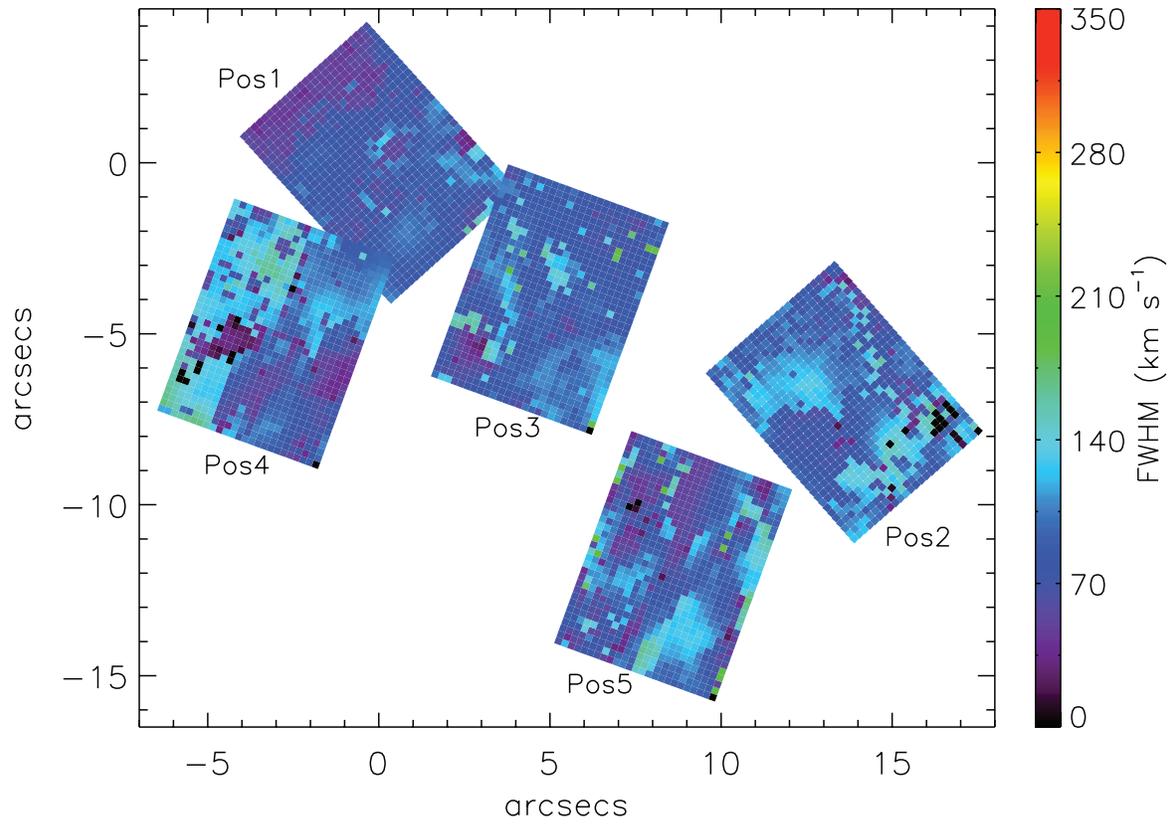} 
\caption{H$\alpha$ FWHM maps for C1. A scale bar is given for each map in \kms\ units, corrected for instrumental broadening. \textit{[A colour version of this figure is included in the on-line version.]} }
\label{fig:Hac1_fwhm}
\end{figure*}

\begin{figure*}
\centering
\plotone{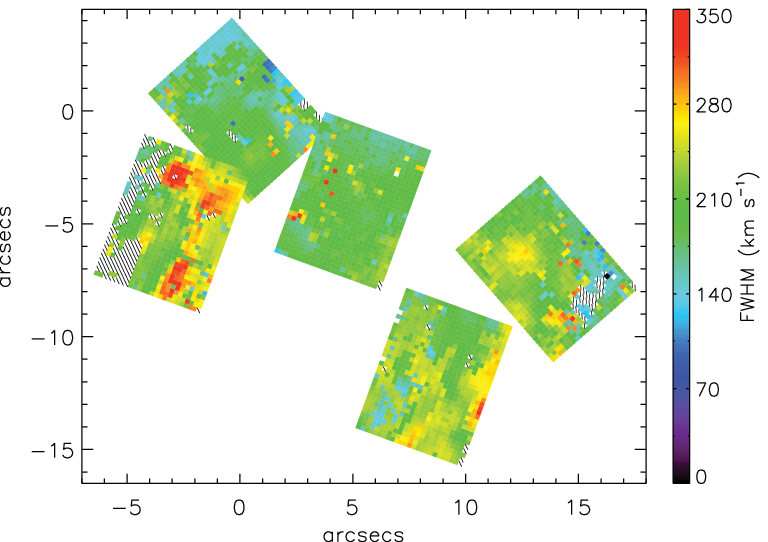} 
\caption{H$\alpha$ FWHM maps for C2. Labels as for Fig.~\ref{fig:Hac1_fwhm}. \textit{[A colour version of this figure is included in the on-line version.]} }
\label{fig:Hac2_fwhm}
\end{figure*}

\begin{figure*}
\centering
\plotone{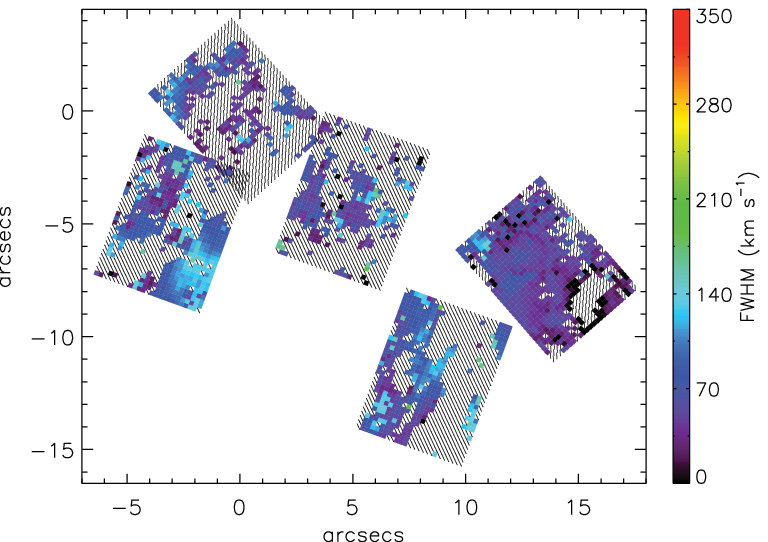} 
\caption{H$\alpha$ FWHM maps for C3. Labels as for Fig.~\ref{fig:Hac1_fwhm}. \textit{[A colour version of this figure is included in the on-line version.]} }
\label{fig:Hac3_fwhm}
\end{figure*}

Maps of the FWHM of each of the H$\alpha$ line components are shown in Figs.~\ref{fig:Hac1_fwhm}, \ref{fig:Hac2_fwhm} and \ref{fig:Hac3_fwhm}. As mentioned in Section~\ref{sect:line_profiles}, the H$\alpha$ line profile across the five fields is, in general, a convolution of a bright, narrow component (C1) and a broad, fainter component (C2). The majority of position 1, for example, exhibits this kind of profile. In position 4 (and some of position 3), however, we see a split narrow component (C1 and C3), and in some places a very broad underlying component (up to $\sim$350~\kms). The kinematic structure of these inner regions is clearly complex, with evidence for increasing complexity and irregularity away from the starburst clumps, out into the inner wind. We will now briefly describe some of the details of these maps.

In C1, the line widths are fairly uniform across the five fields at $\sim$30--120~\kms, but increase to $\lesssim$150~\kms\ in a number of localised regions. The widths of C2 (Fig.~\ref{fig:Hac2_fwhm}) are quite uniform across positions 1 and 3 (130--250~\kms). However, there is a distinct peak in the centre of position 2 where the line widths reach $\sim$270~\kms. This is coincident with the flux peak of region C and the `region C bubble' discussed below (Section~\ref{sect:GMOS_vel}). In positions 4 and 5, the FWHM variations trace north-west--south-east oriented patterns, and reach $>$300~\kms\ in places. The widths of C3 (Fig.~\ref{fig:Hac3_fwhm}) are in general similar to or less than the C1 FWHMs; in the south-west of position 4, however, we do see the C3 line widths consistently increasing above 100~\kms, and in position 2 (region C) their widths are narrower than C1. Table~\ref{tbl:GMOS_fwhm} lists the average line widths for the three line components over all five fields (the errors quoted represent the standard deviation on the sample and do not reflect errors on the individual measurements).

In position 4, we find a compact region of very broad-line gas ($\lesssim$350~\kms) near the centre-north, evident in C2 only. Hereafter we refer to this region as the `position 4 knot'. An example line profile from this knot is shown in Fig.~\ref{fig:egfits}d, showing how at this point C2 is not only very broad, but highly blueshifted ($\gtrsim$150~\kms; see below). We discuss its possible origins in Paper~II after analysing the line ratio and excitation information as well as \textit{HST} images. We come to the conclusion that this unresolved source is likely to be a star surrounded by a compact nebula, possibly a Luminous Blue Variable.

\begin{table}
\begin{center}
\caption{Average FWHMs (mean $\pm$ standard deviation) for the three line components over all spaxels in all fields for the GMOS and DensePak data.}
\label{tbl:GMOS_fwhm}
\begin{tabular}{l r @{\,$\pm$\,} l r @{\,$\pm$\,} l}
\hline
Component & \multicolumn{2}{c}{Average GMOS FWHM} & \multicolumn{2}{c}{Average DensePak FWHM}\\
& \multicolumn{2}{c}{(\kms)} & \multicolumn{2}{c}{(\kms)} \\
\hline
C1 & 90 & 30 & 116 & 41 \\
C2 & 212 & 39 & 224 & 52 \\
C3 & 68 & 34 & 103 & 40 \\
\hline
\end{tabular}
\end{center}
\end{table}

In summary, we find uniform line widths in C1 and C2 across regions A and C in positions 1, 2 and 3 (particularly in region A). The width of the broad component (C2) is on average $\sim$200~\kms, rising to over 300~\kms\ in places. In the inner wind regions it traces linear patterns parallel to minor-axis; these are particularly obvious in C2.

\subsection{Radial velocities} \label{sect:GMOS_vel}

\begin{figure*}
\centering
\plotone{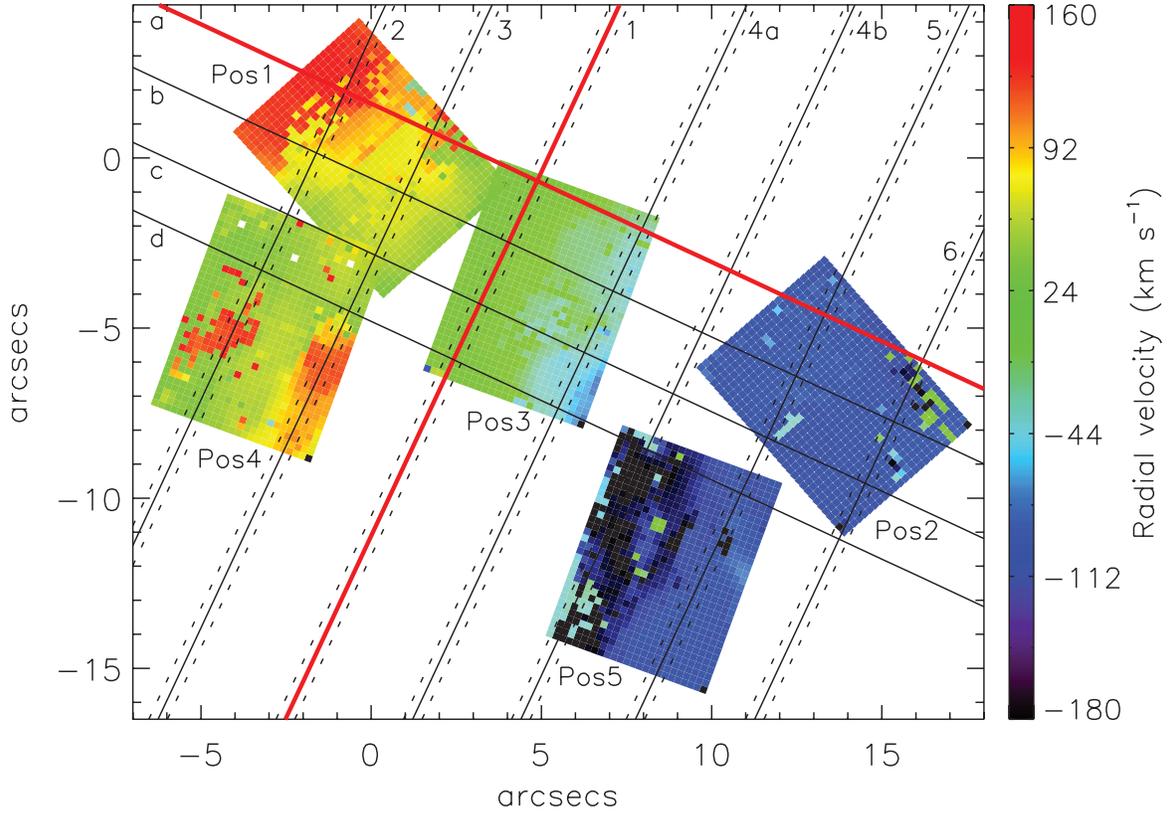} 
\caption{GMOS H$\alpha$ radial velocity map for C1. A scale bar is given in \kms\, relative to $v_{\rm sys}$ (200~\kms). Four major-axis pseudo-slits ($0\farcs4$ width) are represented as \textit{lettered} lines running north-east to south-west: slit a runs along the galaxy major-axis (PA = 65$^{\circ}$; highlighted in red). Major-axis slits b, c and d are parallel to slit a but offset by 2, 4 and 6 arcsecs, respectively. Seven minor-axis pseudo-slits are also indicated by \textit{numbered} lines running north-west to south-east: slit 1 runs along the galaxy minor-axis (PA = 160$^{\circ}$; also highlighted in red); the widths of these minor-axis pseudo-slits are indicated with dashed lines in each case. Minor-axis slits 2, 3, 4a and b, 5 and 6 are parallel to slit 1 but offset by $-6$, $-3.2$, +3.5, +6.0, +9.5 and +12 arcsecs, respectively. The resulting position-velocity diagrams are plotted in Figs.~\ref{fig:GMOS_maj} and \ref{fig:GMOS_min}. \textit{[A colour version of this figure is included in the on-line version.]} }
\label{fig:Hac1_allpos_vel_slits}
\end{figure*}

\begin{figure*}
\centering
\plotone{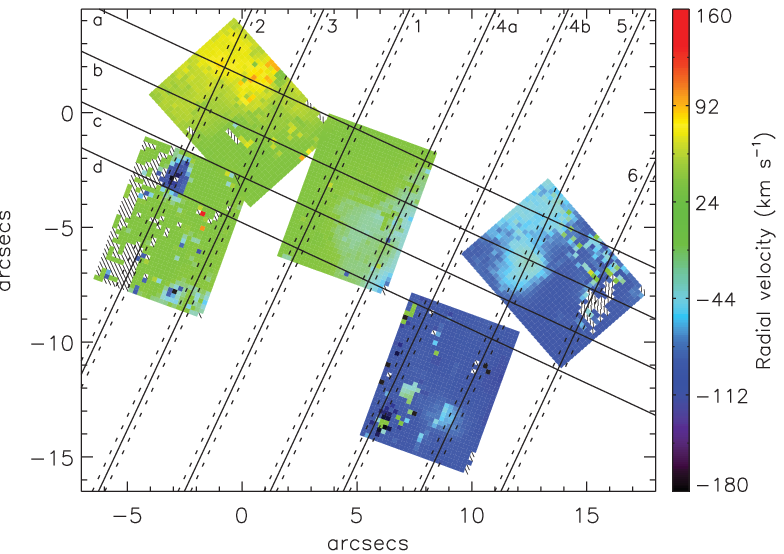} 
\caption{GMOS H$\alpha$ radial velocity map for C2. Labels as for Fig.~\ref{fig:Hac1_allpos_vel_slits}. \textit{[A colour version of this figure is included in the on-line version.]} }
\label{fig:Hac2_allpos_vel_slits}
\end{figure*}

The aforementioned radial velocity maps are presented in Figs~\ref{fig:Hac1_allpos_vel_slits}, \ref{fig:Hac2_allpos_vel_slits} and \ref{fig:Hac3_allpos_vel_slits} for the three H$\alpha$ line components. The highest positive (redshifted) velocities in all three components are found in the north-east of position 1. However on the spaxel-scale there are distinct morphological differences between the three component maps: in C1 there is a smooth gradient in velocity peaking at $\sim$160~\kms\ and falling towards the south-west, whereas in C2 the velocities peak at only $\sim$100~\kms. 
The overall red--blue velocity gradients seen in the three components over the five IFU positions reflect the large-scale rotation of the galaxy (including the bar in the inner $\sim$$20''$). However, overall they do not appear to be aligned with the photometric major axis. For example, the highest negative (blueshifted) velocities are found in position 5 (not 2), and again are much higher in C1 ($-150$ to $-180$~\kms) than C2 ($-100$ to $-150$~\kms). If the gas follows normal disk rotation with an axis aligned with the major axis, then the most blueshifted velocities would be expected to be in position 2 since its coverage extends further to the west. This tilt in the axis of rotation is in agreement with what was found by \citet[][see their fig.~8a]{r-r04} from radio observations of the H92$\alpha$ recombination line, and will be discussed more in Section~\ref{sect:DP_Ha_vel}.

Disregarding the large-scale differences due to galaxy rotation, the C1 and C2 velocities in position 2 (region C), appear more uniform and smooth than in position 1 (region A). Coincident with the peak in H$\alpha$ flux and line widths in the centre of region C, we see a region of C2-emitting gas with distinctly more positive (redshifted) velocities to its surroundings (by $\lesssim$50~\kms). As alluded to above, we will refer to this as the region C bubble.

The position 4 knot (identified above as a compact region of very broad C2-emitting gas near the centre-north of position 4) shows up distinctly in the C2 radial velocity map (Fig.~\ref{fig:Hac2_allpos_vel_slits}). Clearly the knot is distinct (at least kinematically) from its (projected) surroundings. As mentioned above, the nature of this knot is discussed fully in Paper~II.

The existence of a third H$\alpha$ component provides evidence for increased dynamic complexity. In the north of position 1 we see C1 and C3 lines of $\sim$25--35~\kms\ separation. These extend into the centre of region A. We also identify split lines with separations from $\sim$10--75~\kms\ in an area in position 3, towards the west of region A. The location of these split lines trace the H$\alpha$ morphology well. The form of C1 and C3 velocities in positions 4 and 5 (the wind positions) follow linear patterns oriented along the minor axis, as we found for the equivalent FWHM maps. In position 5, we interpret this splitting, in conjunction with line width (see above) and density (Paper~II) measurements and morphologies, as originating from the walls of a distinct filament (channel/chimney) in the outflow. These ideas are introduced in Section~\ref{sect:disc} and discussed more fully in Paper~II. The line separations are $\sim$25~\kms\ in the north-east, 125--150~\kms\ in the west and south-west, and 50~\kms\ in the centre-east. Throughout position 2 (region C), line splitting is also present. In the north-eastern half, the red component is stronger (hence the blue component was assigned to C3), whereas in the south-western half, the blue component is stronger (and the red component was assigned to C3); the line separation is $\sim$50--70~\kms.

In summary, we see H$\alpha$ line splitting of $\sim$20--70~\kms\ throughout the starburst clumps A and C. This line splitting increases with radius to over 100~\kms\ in the inner wind positions, and appears to follow minor axis oriented patterns.

\subsubsection{Major axis position-velocity graphs}
Although the type of maps described above are the most complete way of representing this type of 3D dataset, it can still be very difficult to interpret the results in such a fashion: the data need to be `reduced' further. Thus, we have defined a number of pseudo-slits across our fields, from which we can extract position-velocity diagrams.

\begin{figure*}
\centering
\plotone{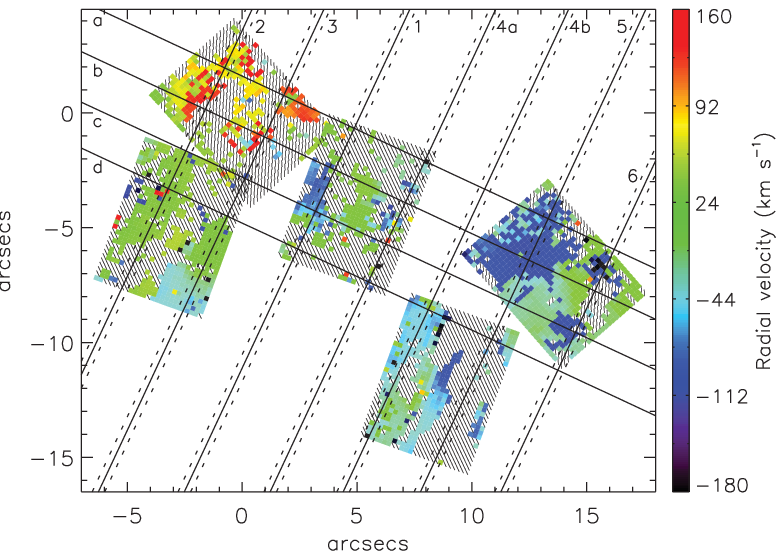} 
\caption{GMOS H$\alpha$ radial velocity map for C3. Labels as for Fig.~\ref{fig:Hac1_allpos_vel_slits}. \textit{[A colour version of this figure is included in the on-line version.]} }
\label{fig:Hac3_allpos_vel_slits}
\end{figure*}

\begin{figure*}
\centering
\plotone{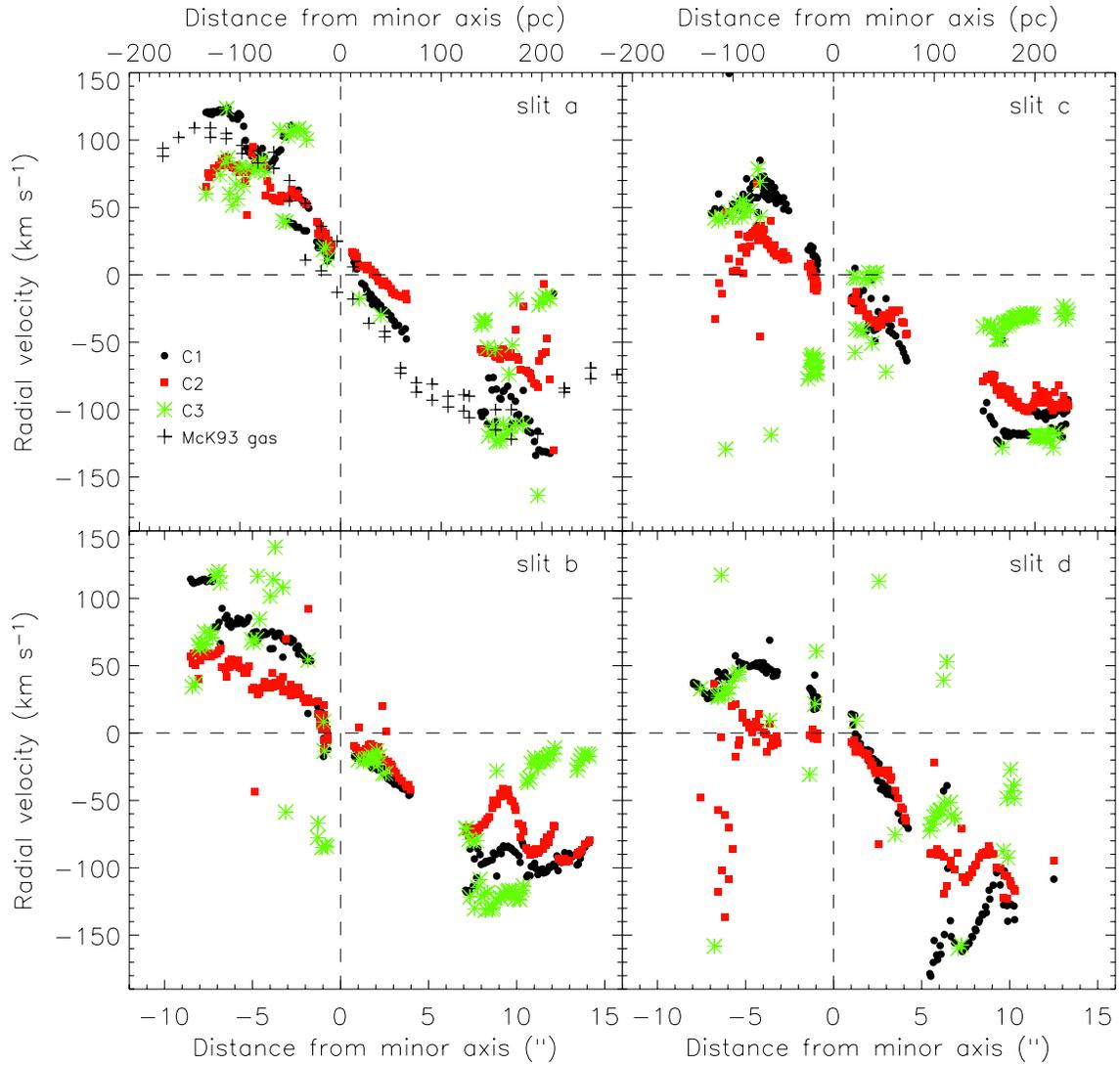} 
\caption{H$\alpha$ position-velocity plots for the four major-axis pseudo-slits shown in Figs~\ref{fig:Hac1_allpos_vel_slits}, \ref{fig:Hac2_allpos_vel_slits} and \ref{fig:Hac3_allpos_vel_slits}. The gaseous ([S\three] and P10) and stellar (Ca\two) radial velocity measurements of \citet{mckeith93} are plotted in the upper-left panel (slit `a') for comparison. Radial velocities are relative to $v_{\rm sys}$, and the $x$-axes show the distance from the minor-axis in both arcsecs (bottom) and parsecs (top; assuming $1'' = 17.5$~pc). The radial velocity error bars are for the most part smaller than the symbol sizes, except for C3 where the uncertainties lie in the range 10--30~\kms\ (see Section~\ref{sect:line_profiles}) but these have been omitted for clarity.} \textit{[A colour version of this figure is included in the on-line version.]}
\label{fig:GMOS_maj}
\end{figure*}

The four \textit{lettered} lines running north-east to south-west on Figs~\ref{fig:Hac1_allpos_vel_slits}, \ref{fig:Hac2_allpos_vel_slits} and \ref{fig:Hac3_allpos_vel_slits} indicate the location of four $0\farcs4$ wide major-axis pseudo-slits from which we have extracted the radial velocities of each H$\alpha$ line component. Slit `a' runs along the galaxy major axis (PA = 65$^{\circ}$); slits `b', `c' and `d' are parallel to slit `a' but offset by 2, 4 and 6 arcsecs (35, 70 and 100~pc) respectively. The resulting position-velocity diagrams are plotted in Fig.~\ref{fig:GMOS_maj}. Near-IR [S\three] and P10 radial velocity measurements from \citetalias{mckeith93} are plotted in the slit `a' graph for comparison (labelled `\citetalias{mckeith93} gas').

A number of observations can immediately be drawn from these plots. Firstly, the velocities of C1 (black points) along the major axis (slit `a') follow the near-IR emission line measurements of \citetalias{mckeith93} (+ symbols) very well. In the central, steepest part (inner 15--20$''$), we measure a velocity gradient in C1 of $\sim$15~\kms~arcsec$^{-1}$, in excellent agreement with previously published measurements \citepalias{mckeith93, shopbell98, westm07c}. This steep gradient is due to the bar $x_{2}$-orbits (\citealt{wills00}; \citetalias{westm07c}). The fact that we can see the turn-over in bar orbit velocities on \textit{both} sides of the nucleus implies that we are seeing emission from the whole bar and thus more than half-way into the galaxy disk. This challenges the canonical idea of a highly obscured nuclear region.

Secondly, the velocity gradient of C2 (red points) is much shallower than C1 (black points) in all 4 pseudo-slits. For example, in the same region of slit `a' as the C1 velocity gradient was measured, we find a C2 velocity gradient of $\sim$9~\kms~arcsec$^{-1}$. The rotation of the gas emitting C2 must therefore be slower than that of the C1-emitting gas within the disk and southern inner-wind. Furthermore, the amplitude of the C1 velocity curve decreases from slits `a' to `c' (clearly seen by comparing the velocities at, for example, $-5''$ offset), although some contribution to this effect may result from the offset between the major axis and the rotation axis of the gas. Slit `d' is dominated by small-scale velocity structure, making it difficult to measure meaningful gradients. However, it is clear that the C2 curve is still considerably flatter than that of C1. We discuss what these findings imply in Section~\ref{sect:disc}.

We now highlight some of the many small-scale dynamical features seen in these plots. At $-4''$ to $-3''$ in slit `a', we see a coherent rise in velocities traced by C1 and C3, indicating the presence of a kinematically distinct gas component. This rise, also observed in the STIS H$\alpha$ observations of \citetalias{westm07c}, occurs right in the core of region A, and is visible as a red wedge-shaped region in the position 1 C3 velocity map (Fig.~\ref{fig:Hac3_allpos_vel_slits}). Its existence suggests moderately fast shell expansion: assuming that the two components represent the two halves of a bubble-like structure, we can use (half) the velocity difference between the two components as an indicator of the approximate expansion speed. Here we measure speeds of $\sim$30~\kms.

Large C1--C3 line splitting is seen between approximately +5 to +15$''$ offsets (region C) in all four slits, with line separations of 80--100~\kms. Interestingly, here the velocity of the broad C2-emitting gas mirrors that of the C1 gas, but always with a redshift. This large line splitting indicates that here we are probing a large column depth into the galaxy. A closed velocity ellipse can clearly be identified centred at +9$''$ in slit `b' (traced by C1 and C3), measuring $\sim$$4''$ (70~pc) in extent. This corresponds to the region C bubble mentioned above. The expansion velocity of this structure is $\sim$20~\kms. Here again the C2 velocities mirror those of C1 but with a distinct redshift. 

These small-scale features cannot simply arise from optical depth/obscuration effects, since we would then expect the velocities to always be blueshifted relative to the local value (i.e.\ arising exclusively from the near side of the wind), which we do not observe. Identifying structures such as these, and the correspondence of the C2 velocities with these features, would be very difficult to identify directly from the individual IFU maps, thus showing the advantages of plotting the data in this `reduced' form.

To summarise, the velocity gradient in C2 follows a distinctly shallower gradient than C1 -- this is evident in all four slits (i.e.\ out to 100~pc from the major axis). The fact that we see the turn-over in the galaxy rotation curve on both sides of the nucleus, and large C1--C3 line splitting at $\sim$+10$''$ on the major axis, implies that we are seeing a considerable distance into the disk. This is in conflict with the high levels of obscuration usually associated with the nuclear regions of M82. Evidence of local, small-scale ($\sim$100~pc size) line splitting is found, implying structures expanding at 20--50~\kms, including one clear coherent bubble in region C. Even though C2 has a different large-scale gradient to C1, it follows the local velocity patterns of the narrow components, indicating that the two components may be dynamically connected.

\subsubsection{Minor axis position-velocity graphs}
\begin{figure*}
\centering
\plotone{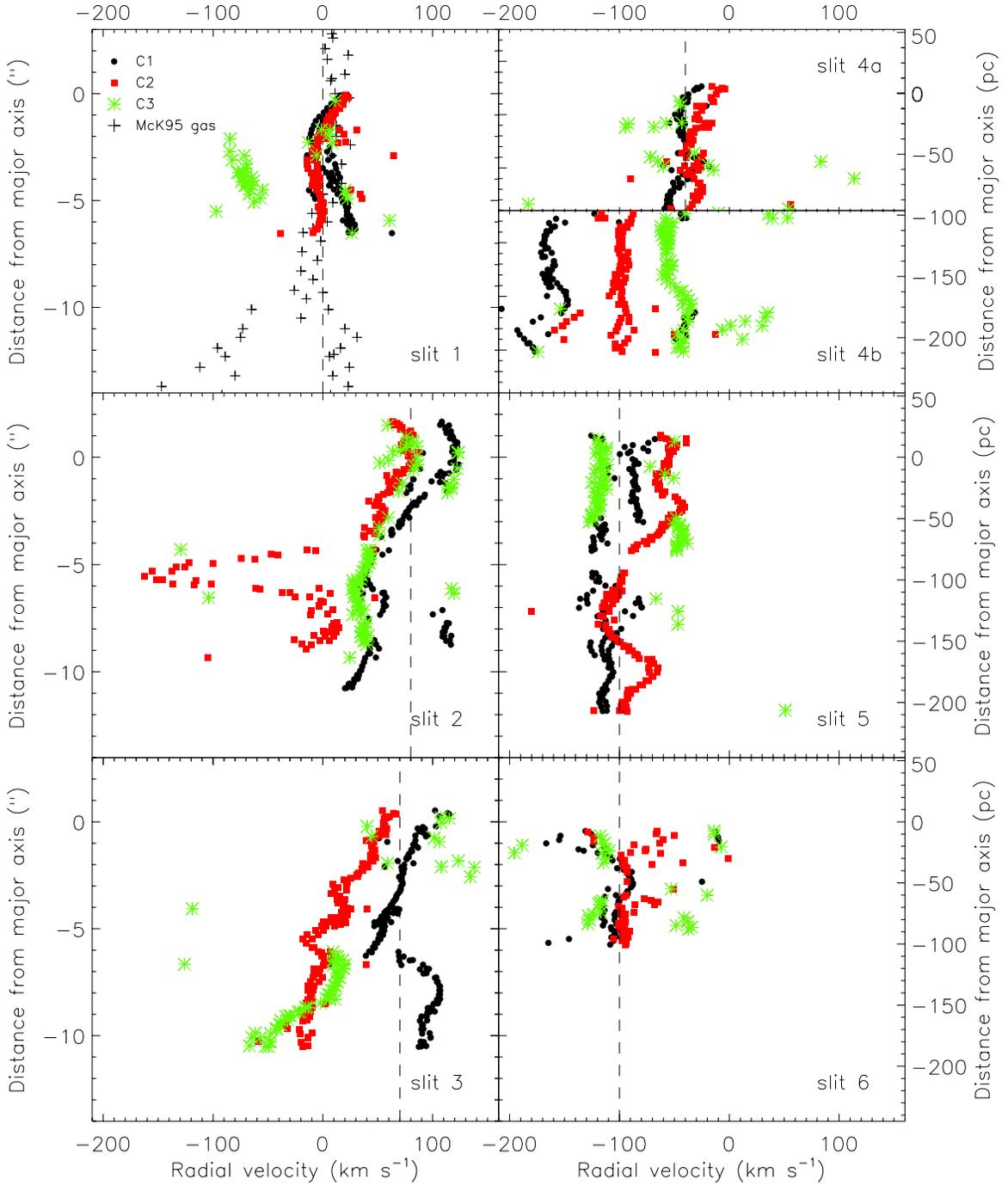} 
\caption{H$\alpha$ position-velocity plots for the six minor-axis pseudo-slits shown in Figs~\ref{fig:Hac1_allpos_vel_slits}, \ref{fig:Hac2_allpos_vel_slits} and \ref{fig:Hac3_allpos_vel_slits}. The H$\alpha$, [N\two]$\lambda$6583, and [S\three]$\lambda$9532 emission line radial velocity measurements of \citetalias{mckeith95} are plotted in the upper-left panel (slit 1) for comparison. Radial velocities are relative to $v_{\rm sys}$, and the $y$-axes show the distance from the major-axis in both arcsecs (left) and parsecs (right; assuming $1'' = 17.5$~pc). The vertical dashed lines represent the approximate radial velocity of the gas at the point at which the pseudo-slit crosses the major axis, thus showing the zero point with respect to the galaxy rotation. Again, the radial velocity error bars are for the most part smaller than the symbol sizes, except for C3 where the uncertainties lie in the range 10--30~\kms\ (see Section~\ref{sect:line_profiles}) but these have been omitted for clarity. \textit{[A colour version of this figure is included in the on-line version.]} }
\label{fig:GMOS_min}
\end{figure*}

To help us interpret the evolution of the dynamics in the wind direction, we have also defined seven minor-axis pseudo-slits; these are indicated on Figs.~\ref{fig:Hac1_allpos_vel_slits}, \ref{fig:Hac2_allpos_vel_slits} and \ref{fig:Hac3_allpos_vel_slits} by the numbered lines running north-west to south-east. Slit 1 runs along the galaxy minor axis (PA = 155$^{\circ}$); slits 2--6 are parallel to slit 1 but offset by $-6$, $-3.2$, +3.5, +6.0, +9.5 and +12 arcsecs ($-105$, $-55$, +60, +105, +165 and +210~pc) respectively. We have split slit 4 into `a' and `b' since they pass through the far-western edge of position 3 and the far-eastern edge of position 5. The widths of the pseudo-slits ($\sim$$0\farcs4$) are indicated with dashed lines in each case; the resulting position-velocity diagrams are plotted in Fig.~\ref{fig:GMOS_min}. In the top-left panel, we have included the H$\alpha$, [N\two]$\lambda$6583 and [S\three]$\lambda$9532 minor axis radial velocity measurements from \citetalias{mckeith95} for comparison.

Our velocity measurements along the minor axis (pseudo-slit 1) only extend $\sim$120~pc south of the nucleus. Here the velocities of C1 and C2 are very consistent with one another, and are in good agreement with the \citetalias{mckeith95} data. Between 2--6$''$ from the major axis, we see C1--C3 velocity separations of $\sim$80~\kms. With reference to the velocity map of Fig.~\ref{fig:Hac3_allpos_vel_slits}, we see that here the slit passes through one side of the region containing split lines in position 3. Slit 2 is positioned at the eastern-most extent of our IFU coverage. Here the three components are all redshifted by $\lesssim$+100~\kms\ due to the rotation of the galaxy (indicated by the vertical dashed line), and the lag of C2 behind C1 is clearly reproduced along the entire extent of this pseudo-slit. The most prominent feature is the `knot' of highly blueshifted (up to $-180$~\kms) C2-emitting gas located at $-6''$ ($-95$~pc) from the major axis. In slit 3, we see a gradual divergence of the C1 and C3 velocities from $-7''$ downwards; at $-11''$ the two components are separated by $\gtrsim$160~\kms. This gradual rise is mirrored by C2, where at $-11''$ the C1--C2 velocity separation is $\sim$120~\kms.

As described above, pseudo-slit 4 was split into two since positions 3 and 5 are positioned close to one another but are not spatially contiguous. The discontinuity is represented in Fig.~\ref{fig:GMOS_min} by splitting the plot into two. In slit 4a (position 3), the C1 and C2 velocities are separated by only 20--30~\kms\ (where C2 is redder than C1). In position 5, however, just a few arcsecs away, we can identify a blueshifted and brighter narrow component (black C1 points at $-150$ to $-180$~\kms). We interpret the line emission referred to as C1 in slit 4a (position 3) to be C3 here in slit 4b (since in assigning the line components we define C3 to be the \textit{faintest} of any narrow lines identified). This highly blueshifted emission is not present in position 3, but identified throughout position 5, indicating that we may miss a significant transition region between the two IFU fields. At $-14''$ from the major axis in slit 4b, the separation of C1 and C2 is $\sim$70~\kms\ and of C1 and C3 is $\sim$130~\kms. 

In contrast, the minor axis velocity patterns in slit 5 are quite consistent between positions 2 and 5 (despite the small gap in spatial coverage). On this side of the galaxy, the velocity of the C2-emitting gas again can clearly be seen to lag behind the C1-emitting gas (now in the opposite sense). In position 2 (small offsets), C3 is found on both sides of C1; separations reach up to 40~\kms\ for the blueshifted C3 and 60~\kms\ for the redshifted C3 emission. The pseudo-slit was purposefully positioned to pass through a part of IFU position 5 that contains C3 emission, and we find C3 to be redshifted with respect to C1 by $\lesssim$70~\kms. The spatial coverage of slit 6 is fairly restricted; this pseudo-slit is located at the western-most edge of our IFU coverage, and as such, all the velocities are blueshifted by $\sim$$-100$~\kms\ due to galaxy rotation (again highlighted by the vertical dashed line). The presence of high velocity C3 demonstrates the presence of disturbed kinematics in this part of the disk.

As mentioned above, our spatial coverage of the minor axis is very limited; we will therefore make use of the \citetalias{mckeith95} minor axis data in order to compare the results from the different slits. At $\gtrsim$$-9''$ offset, \citetalias{mckeith95} identify two rapidly diverging line components, and interpret them as being emitted from the front and back walls of the large-scale wind `cone', where the increasing velocity separation is taken to be evidence of a rapidly accelerating outflow. This diverging pattern is clearly mirrored in slit 3 (only $\pm$$\sim$70~pc from the minor axis), where we can equate their ``red and blue'' line components to our C1 and C3. In this slit, however, the radius at which C1 and C3 begins to separate appears to be much smaller ($\lesssim$$7''$). In slits 2 and 5, the situation is reversed, and we see C1--C3 line splitting at small radii (of $\sim$40~\kms). These results attest to the existence of a complex kinematic environment here at the base of the large-scale outflow cone, with many overlapping expanding structures located at different positions.

Where we observe narrow-line splitting, we find that, in general, the strongest component (C1) on the eastern side is the redder one, whereas on the other side of the disk, the strongest component is the bluer one. Thus in the outflow cone interpretation, the back wall of the cone is brightest on the eastern side (which is moving away from us due to the galaxy rotation), whereas the front wall is brightest on the western side. Furthermore, all six minor axis pseudo-slits show evidence for increasingly blueshifted line emission with distance south of the major axis. This is consistent with previous observations of the minor axis velocities (\citetalias{mckeith95, shopbell98}; \citealt{greve04}), and is due to the inclination angle of the galactic disk (oriented such that the southern outflow is directed toward the observer).

\subsubsection{Summary: GMOS radial velocities}
There is a great deal of information contained in the GMOS radial velocity maps for the five IFU positions and three line components. Here we briefly summarise our main findings from these data. Firstly, we find that the (major axis) velocity gradient in C2 follows a distinctly shallower gradient than C1, and this is evident at distances of up to 100~pc from the major axis. However, even though C2 exhibits this large-scale gradient difference, it clearly follows the local velocity patterns of the narrow components, indicating that the two components are somehow dynamically connected. We also find that the amplitude of the C1 rotation curve decreases with distance from the disk midplane.

Evidence of H$\alpha$ line splitting (i.e.\ presence of C3) is found throughout the five regions observed. In clumps A and C, its amplitude is $\sim$20--70~\kms; this rapidly increases to $>$100~\kms\ in the inner wind, where the morphology of the split line regions follows radially oriented patterns. In the outflow cone interpretation (\citealt{gotz90}; \citetalias{mckeith95}), this increase reflects an accelerating outflow, where the line emission originates from the front and back walls of the flow. Local, small-scale ($\sim$100 pc size) coherent features in the C1 and C3 lines are interpreted as evidence of structures expanding at 20--50~\kms\ within this outflow. Now that we can identify and map each individual line component, it is clear that this central region, at the base of the large-scale outflow cone, represents a chaotic, complex kinematic environment with many overlapping expanding structures located at different radii. The interpretation of the M82 wind originating from a singular, organised ``outflow cone base'' appears to be a significant over-simplification.

\subsection{FWHM vs.\ radial velocity}
\begin{figure*}
\centering
\plotone{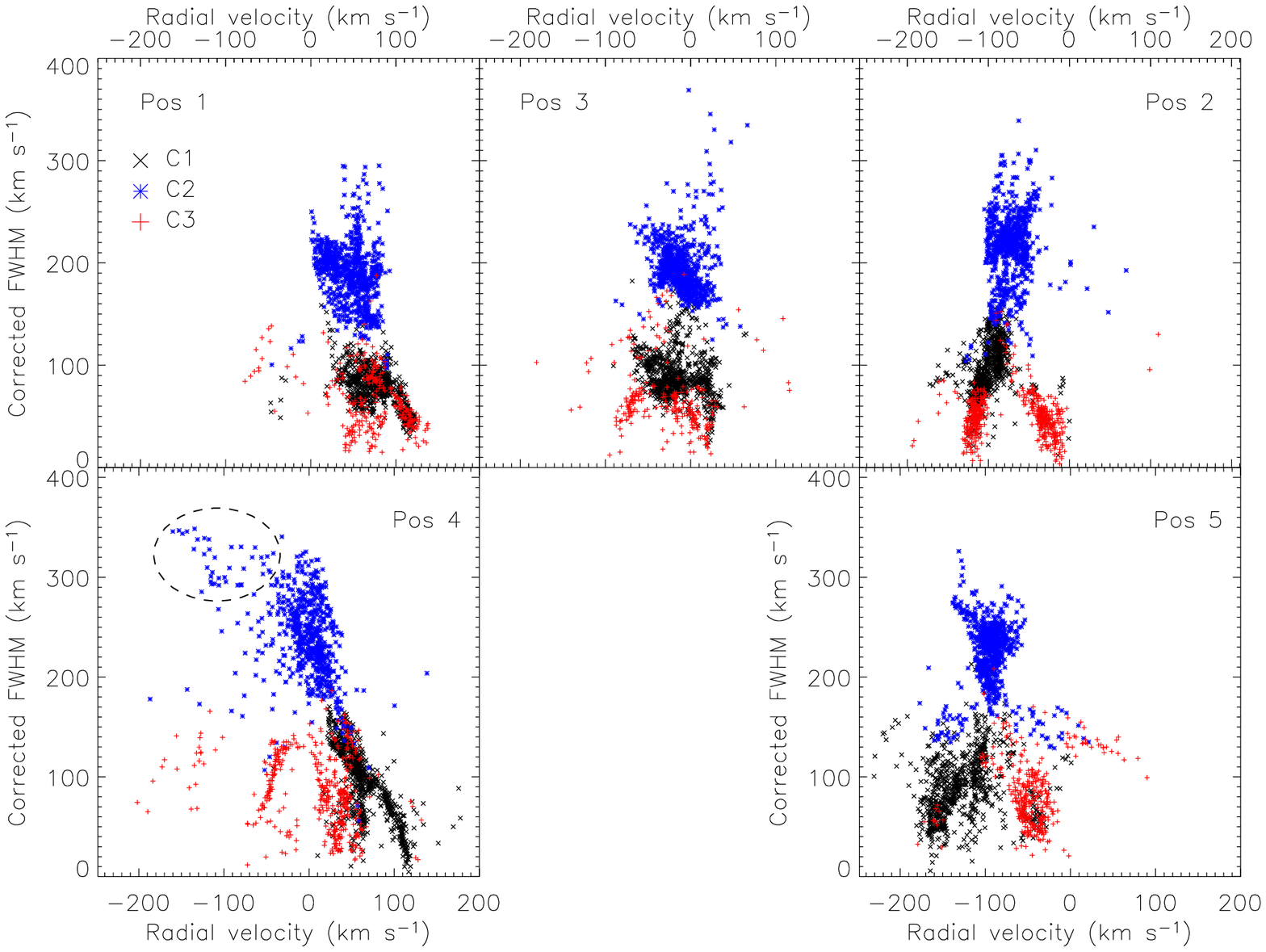} 
\caption{H$\alpha$ FWHM (corrected for instrumental broadening) vs.\ radial velocity (relative to $v_{\rm sys}$) for each component in each IFU field. The plots are displayed in approximately the same order as placed on the sky. The dashed ellipse in the position 4 plot encloses points associated with the `knot' discussed in the text. For an estimate of the associated uncertainties, see Section~\ref{sect:line_profiles}. \textit{[A colour version of this figure is included in the on-line version.]}}
\label{fig:Ha_sigma_vel}
\end{figure*}

An additional way of `reducing' the data further to aid interpretation of the results is to plot the H$\alpha$ FWHM vs.\ radial velocity for each IFU field, as shown in Fig.~\ref{fig:Ha_sigma_vel}.

Plotting the data in this fashion clearly shows the distributions in width- and velocity-space for each of the components in each of the fields, hence illustrating the distinction between C1 (and C3) and C2. Within each field there is little or no systematic offset in velocity between the narrow (C1 and C3) and broad (C2) lines. This is very similar to what we found in NGC 1569 \citep{westm07a, westm07b}. The exception is position 4, where the C1 lines are consistently redshifted compared to both the C2 and C3 points. This position also contains the position 4 knot mentioned above. Points associated with this knot are enclosed by a dashed ellipse in Fig.~\ref{fig:Ha_sigma_vel}.

In positions 1 and 3, C3 lies very much in the same velocity range as C1. In position 2, however, there appears to be two populations of C3 lines, one with similar radial velocities to C1 and one that lies to the redward side. In positions 4 and 5, the population of C3 points lie distinctly to the blueward and redward sides, respectively, of the bulk of the C1 lines. This means that on the eastern side (position 4) the redshifted narrow component is stronger (hence assigned to C1), but on the western side (position 5) the blueshifted component is stronger, reiterating what we found above.

\section{Dynamics of the disk} \label{sect:DP}

In this section we present and describe the 2D FWHM and radial velocity maps derived from our DensePak observations. In contrast to the small-scale, high-resolution GMOS-IFU observations that focus on just the nuclear regions, these data cover most of the inner $\sim$2~kpc of the M82 disk out to disk heights of $>$500~pc (as shown in Fig.~\ref{fig:DPfinder}), albeit at considerably lower spatial resolution. The spectral resolution and S/N of the two datasets, however, are comparable, thus allowing us to make meaningful comparisons between the small- and large-scales as sampled by the two spatially overlapping datasets.

As described in Section~\ref{sect:line_profiles}, the shapes of the emission line profiles throughout much of the four DensePak fields are a convolution of a bright, narrow component (C1) and a broad, fainter component (C2). In some regions, however, we can identify two narrow components, often both superimposed on the broad component. We refer to these components as C1 and C3, and define C3 as the fainter of the two regardless of their wavelength relationship (Section~\ref{sect:line_profiles}). In some places we can identify an H$\alpha$ absorption component (of stellar origin), and fit it using a single Gaussian component (see, e.g., Fig.~\ref{fig:dp_fits}d). To measure the velocity of the Ca\two\ triplet absorption lines we used the pPXF method (again see Section~\ref{sect:line_profiles}).

\subsection{FWHMs} \label{sect:DP_FWHM}

\begin{figure*}
\centering
\includegraphics[width=16cm]{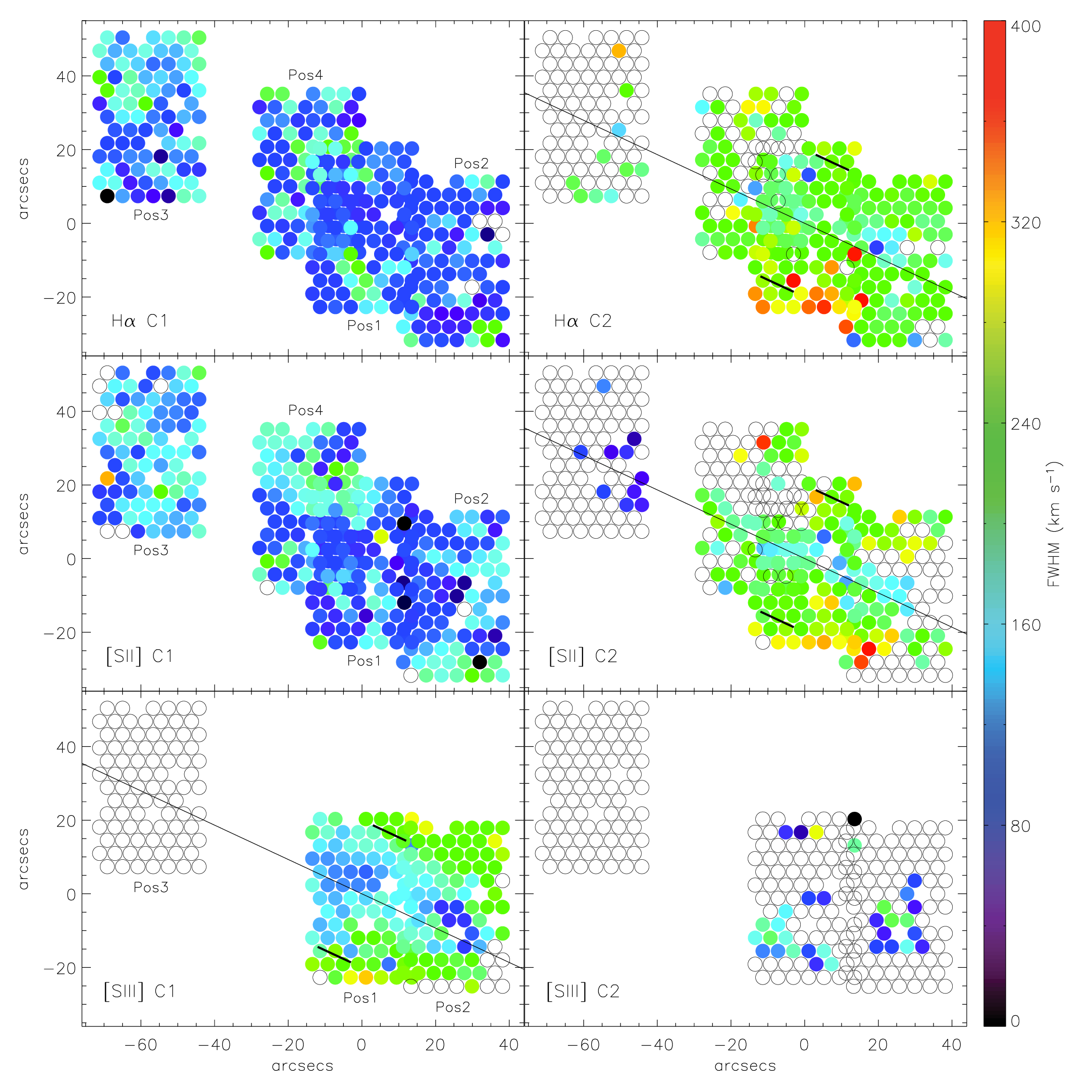} 
\caption{FWHM (corrected for instrumental contribution) maps for the H$\alpha$ C1 and C2 emission components, [S\two]$\lambda$6717,6731 and [S\three]$\lambda$9531. The major axis (PA = $65^{\circ}$) is indicated with a solid line on the H$\alpha$ C2, [S\two] C2 and [S\three] C1 plots. The locations of the two minor axis position-velocity inflection points at $\pm$$20''$ \citep{mckeith95} and discussed in the text are marked with short thick lines. \textit{[A colour version of this figure is included in the on-line version.]} }
\label{fig:dp_fwhm}
\end{figure*}

Maps of the Gaussian FWHM (corrected for instrumental broadening) of the H$\alpha$ C1 and C2 emission components, [S\two]$\lambda$6717,6731 and [S\three]$\lambda$9531 are shown in Fig.~\ref{fig:dp_fwhm}. The first thing to note is that the H$\alpha$ and [S\two] maps are very similar, thus providing a certain level of confidence in our results. The FWHM of C1 of H$\alpha$ and [S\two]\footnote{Recall that the [S\two]$\lambda$6717,6731 doublet were each fit with multiple Gaussian components where the widths of each line were constrained such that FWHM([S\two]$\lambda$6717) = FWHM([S\two]$\lambda$6731)} ranges between 60--170~\kms. The widths of H$\alpha$ C3 (not shown) also fall in this range. In many cases, the spaxels exhibiting broader than average H$\alpha$ (or [S\two]) C1 widths are also those that do not have a corresponding C2 detection. This suggests that in these spaxels C2 does in fact exist, but at too low a level to have been confidently fit, and its presence biases the C1 fits resulting in artificially elevated FWHMs. Where it is detected, H$\alpha$ (and [S\two]) C2 has widths in the range $\sim$200--300~\kms, with the narrowest lines found near the major axis. Where we have common spatial coverage with our GMOS data, our width measurements are in very good agreement. At distances of 400--500~pc south of regions A and C (in positions 1 and 2), the FWHM of H$\alpha$ and [S\two] C2 reaches $\lesssim$370~\kms. This region is consistent with the location of the southern inflection point in the wind outflow velocities identified by \citet{mckeith95}. These infection points are indicated on Fig.~\ref{fig:dp_fwhm} with short thick lines. A similar increase in widths in the north is not convincingly seen, although taking the H$\alpha$ and [S\two] maps together, the line widths certainly do increase away from the disk towards the north.

\begin{figure*}
\centering
\plotone{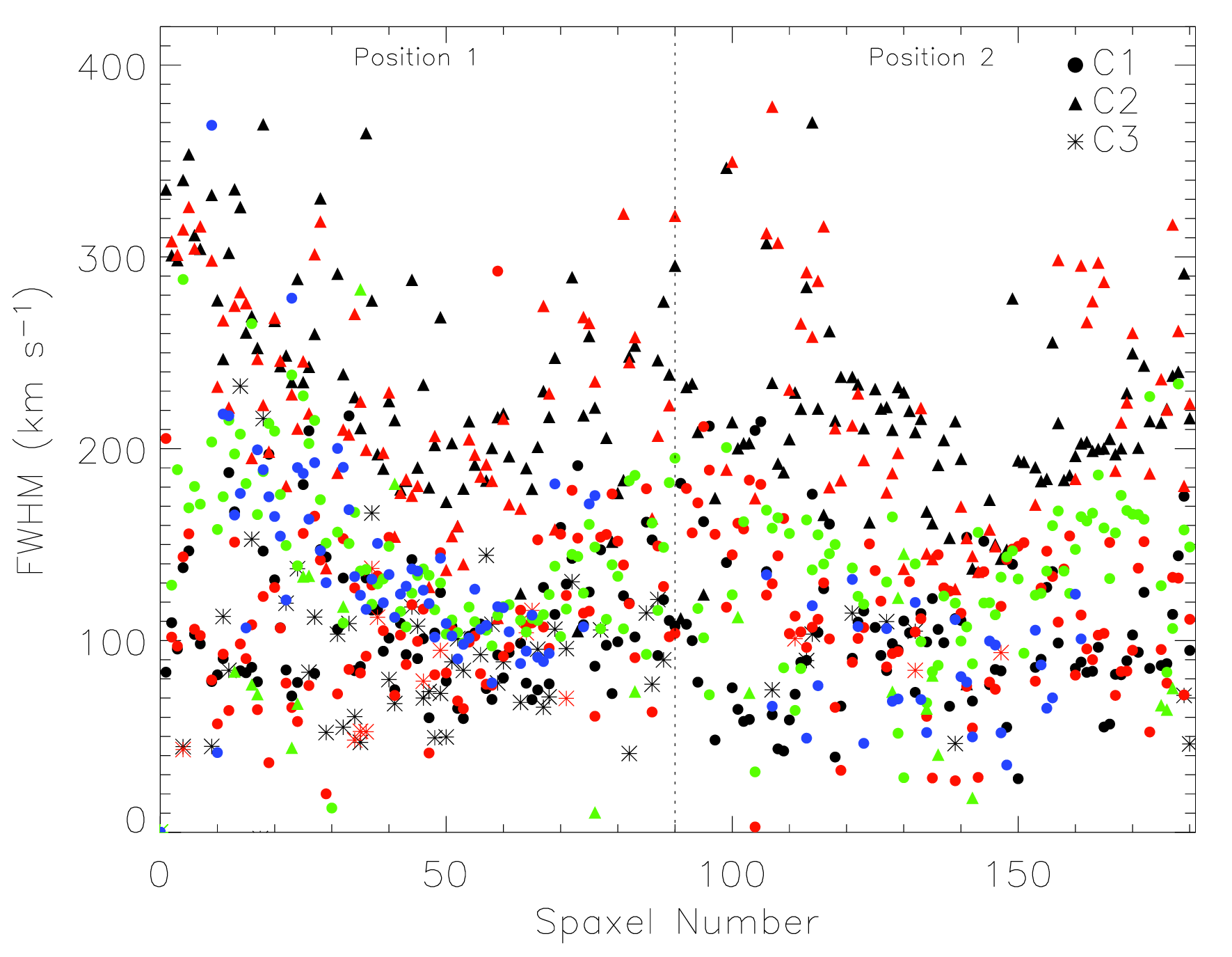} 
\caption{Graph showing the FWHM measurements for positions 1 and 2 (plotted against spaxel number), comparing the widths of all the identified components of H$\alpha$ (black), [S\two] (red), [S\three] (green) and P9 (blue). Although the FWHMs of H$\alpha$ match the equivalent [S\two] widths well (for all three components), the widths of [S\three] C1 are, on average, broader. \textit{[A colour version of this figure is included in the on-line version.]} }
\label{fig:fwhm_compare}
\end{figure*}

The [S\three]$\lambda$9531 FWHM maps (bottom two panels in Fig.~\ref{fig:dp_fwhm}) are distinctly different to the H$\alpha$ and [S\two] maps. Overall [S\three] C1 is significantly broader than the corresponding H$\alpha$ or [S\two] C1. With widths in the range 100--250~\kms\ and the narrowest lines found along the major axis; the [S\three] C1 map in fact has more in common with the H$\alpha$ and [S\two] C2 maps. In Fig.~\ref{fig:fwhm_compare} we compare the FWHMs of all the emission components identified in H$\alpha$, [S\two], [S\three], and P9 for IFU positions 1 and 2. In this representation, the broader widths of the [S\three] C1 lines compared to the equivalent H$\alpha$ (and [S\two]) C1 lines is obvious (in fact the mean H$\alpha$ C1 FWHM in positions 1 and 2 is 99~\kms, whereas the equivalent [S\three] C1 average is 127~\kms).

\begin{figure*}
\centering
\begin{minipage}{7cm} 
\includegraphics[width=7cm]{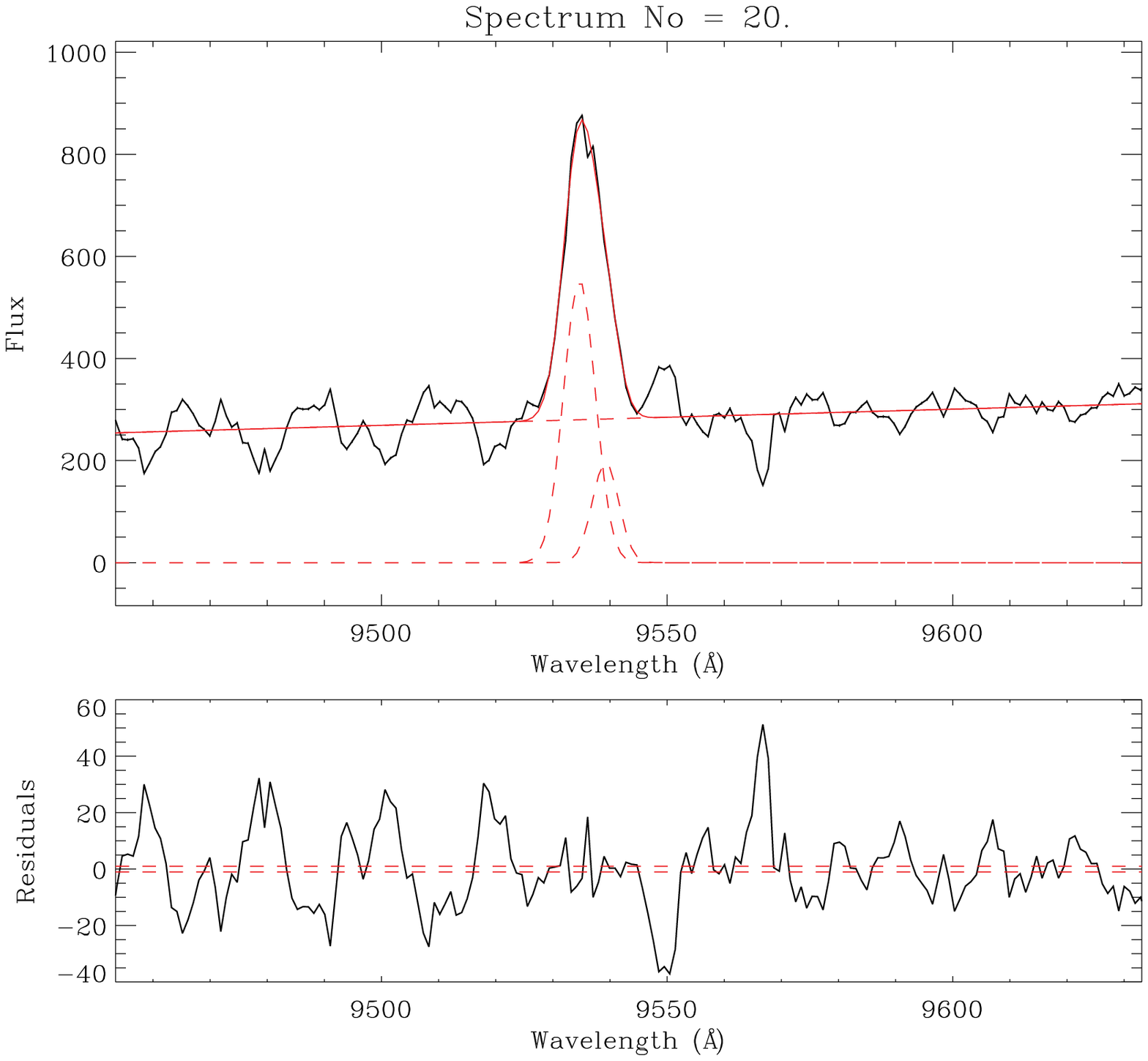} 
\end{minipage}
\hspace{0.2cm}
\begin{minipage}{7.2cm}
\includegraphics[width=7.2cm]{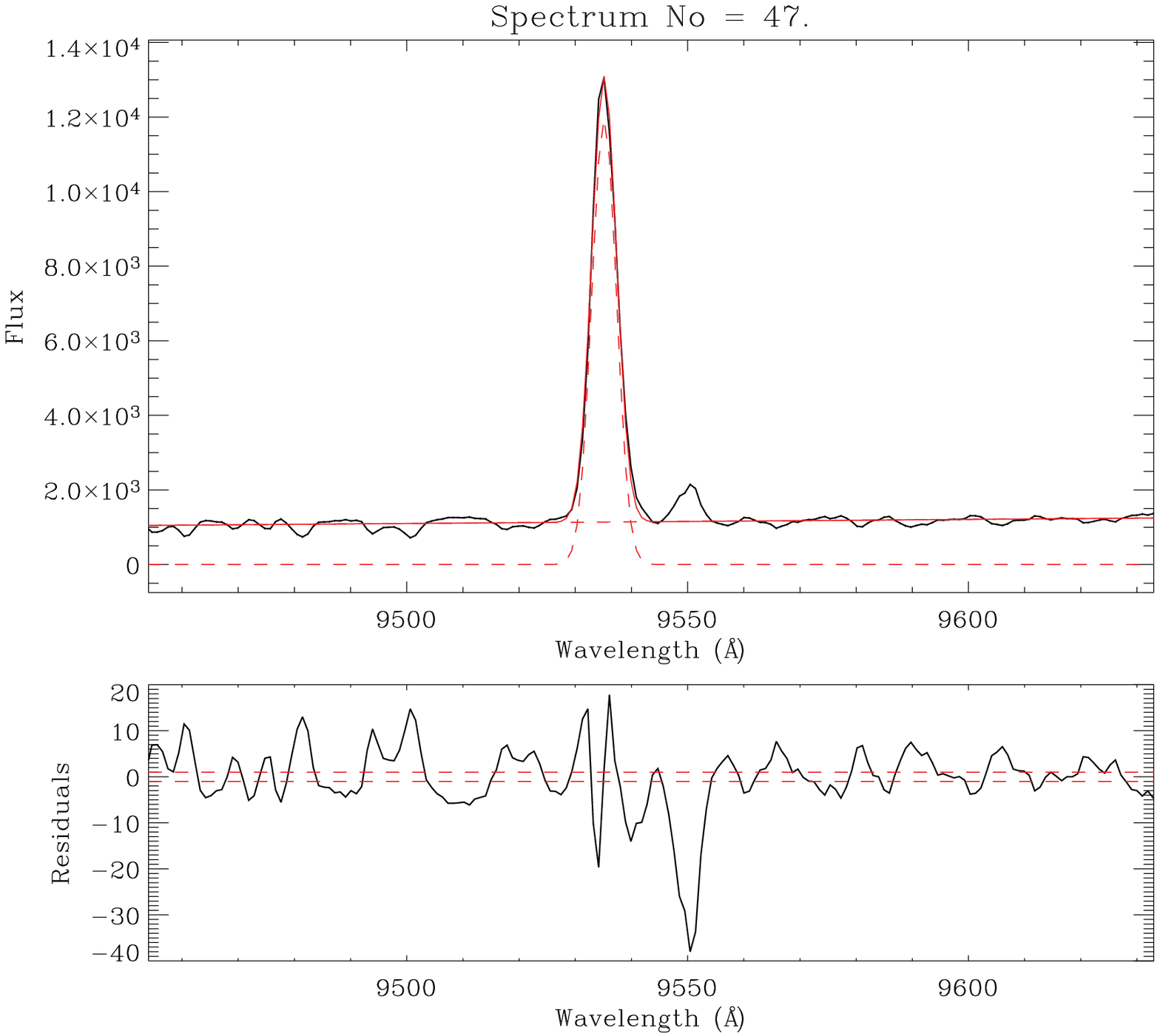} 
\end{minipage}
\caption{Example [S\three]$\lambda$9531 line profiles and Gaussian fits from the south  (left panel) and centre (right panel) of position 1. The C1 FWHMs of each profile are 205 and 160~\kms, respectively, and there is no evidence for broad underlying emission. The lower panels show the fit residuals. }
\label{fig:SIII_profile}
\end{figure*}

The elevated [S\three] C1 widths could result in three ways: (1) [S\three] is probing deeper within the galaxy where the C1-emitting gas is more turbulent ([S\three] is emitted in the near-IR so is less affected by dust extinction); (2) [S\three] is probing a hotter gas phase (S$^{2+}$ has a higher excitation energy than H$\alpha$ or S$^{+}$); or (3) the presence of an unidentified broad component has biased the [S\three] C1 fits to artificially high widths (in a similar way to that described above). To investigate which is the most likely cause, we have performed a number of tests. If the increased widths of the [S\three] line are due to a depth effect (i.e.\ explanation 1), then the near-IR Paschen lines would also exhibit increased widths compared to H$\alpha$. Fig.~\ref{fig:fwhm_compare} shows that this is not the case, and the mean P9 FWHM over positions 1 and 2 is 102~\kms, in good agreement with that of H$\alpha$. Furthermore, the major axis radial velocity gradients of the [S\three] C1 and H$\alpha$ C1 lines are very similar, supporting the conclusion that they both originate at approximately the same depth within the disk. Could the broadening be due to biassed fits? In Fig.~\ref{fig:SIII_profile} we show two example [S\three]$\lambda$9531 line profiles together with the Gaussian models needed to fit their shapes, one originating far from the disk and exhibiting two separated components (left panel), and one high S/N line from near the major axis (right panel). Comparing these to the example line profiles shown in Fig.~\ref{fig:egfits}, it is clear that the near-IR [S\three] lines do not share the same `bright narrow and broad underlying' shapes as the optical emission lines. Even at high S/N, there is no evidence for broad underlying emission, and the profile is well-fit by a single Gaussian. It appears, then, that the [S\three] lines are intrinsically broader than their H$\alpha$ equivalents, and there is no undetected broad emission affecting our fits. We can thus rule out explanation 3. This leaves us with explanation 2, that [S\three] is probing a hotter, more turbulent gas phase than H$\alpha$ C1. We discuss the consequences of this is Section~\ref{sect:disc}.

\subsubsection{Summary: DensePak FWHM results}
In H$\alpha$ and [S\two], the narrowest C2 lines are found near the disk midplane, with the line widths increasing with height above the disk. The broadest C2 widths are found near southern outflow inflection point found by \citetalias{mckeith95}, but there is not an equivalent in the north. The [S\three] C1 widths are much larger than the equivalent H$\alpha$ C1 widths (100--250~\kms\ compared to 60--170~\kms), and the distribution of [S\three] C1 widths closely matches that of H$\alpha$ C2 (i.e.\ narrowest along major axis). After testing a number of possible reasons for this, we conclude that the [S\three] emission arises from a hotter, more turbulent gas phase than that responsible for the H$\alpha$ C1 emission, but, unlike H$\alpha$ C2, originates from the same depth along our line-of-sight through the galaxy as H$\alpha$ C1.

\subsection{Radial velocities} \label{sect:DP_vel}

\subsubsection{H$\alpha$ radial velocities} \label{sect:DP_Ha_vel}

\begin{figure*}
\centering
\plotone{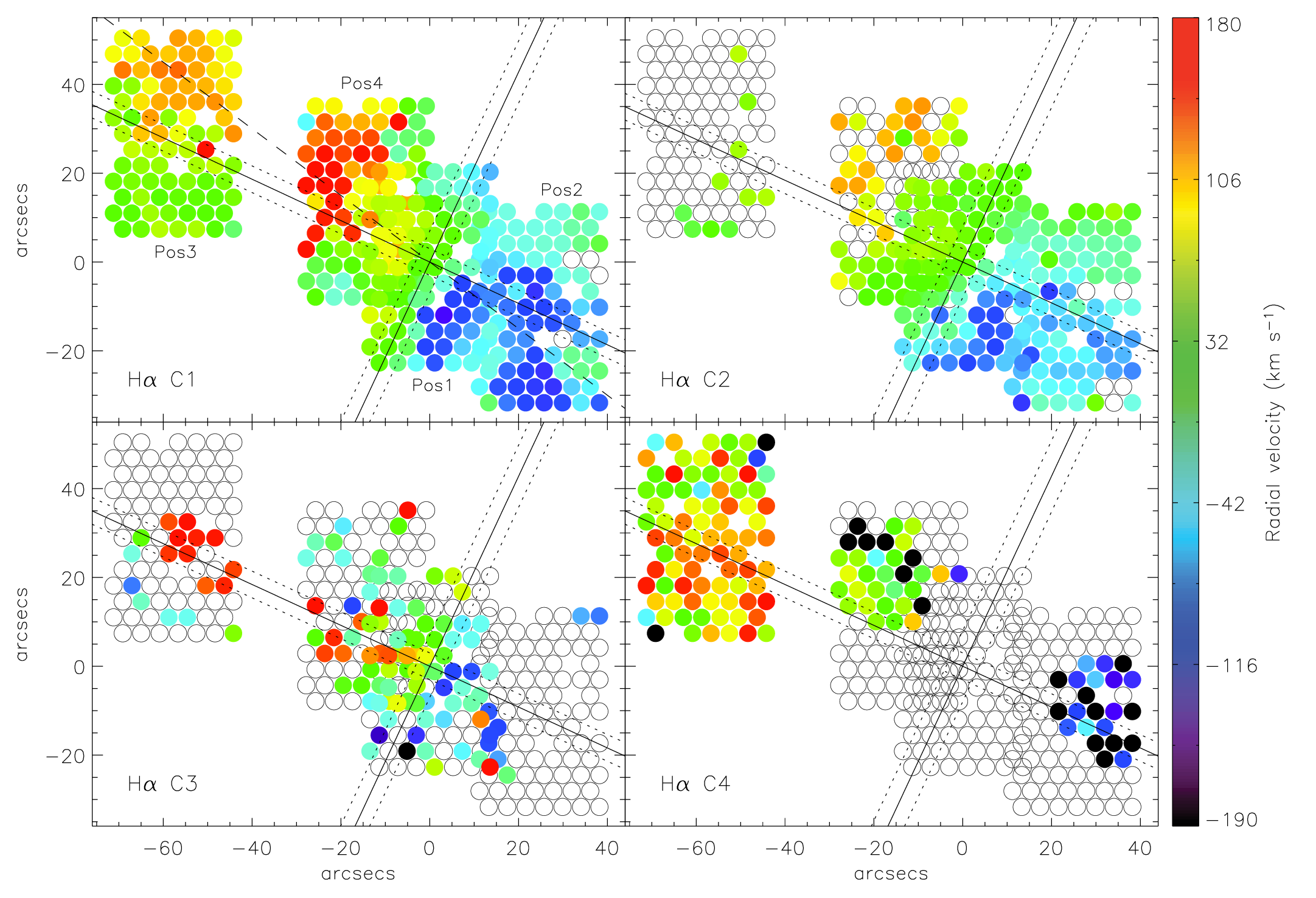} 
\caption{DensePak radial velocity maps for all four H$\alpha$ components (emission: C1, C2, C3, and absorption: C4). The scale is in units of heliocentric \kms\ relative to $v_{\rm sys}$. The galaxy major and minor axes are marked with solid lines, and the dashed lines indicate the width of the pseudo-slits defined in order to make Figs.~\ref{fig:dp_major_axis} and \ref{fig:dp_minor_axis}. The dashed line in the top-left panel represents the PA of the gaseous rotation axis (determined by eye). \textit{[A colour version of this figure is included in the on-line version.]} }
\label{fig:dp_radvel_Ha}
\end{figure*}

Radial velocity maps for the four H$\alpha$ line components (three emission and one absorption) are shown in Fig.~\ref{fig:dp_radvel_Ha}. The large-scale velocity gradient due to galaxy rotation is immediately obvious in the C1 and C2 maps. The highest positive (redshifted) velocities are found in C1 in the east and north-east of position 4 ($\sim$150~\kms), and highest blueshifted velocities of $\sim$$-120$~\kms\ are found in the south-west of position 1, extending into the centre and south of position 2. C2, where detected, closely follows the pattern of C1 velocities, but with a much shallower velocity gradient. This mirrors exactly what we found with GMOS for the nuclear regions, meaning that the C2-emitting gas in the entire disk is rotating at a slower rate than C1. The significance of these findings is discussed in Section~\ref{sect:disc}.

Although the C3 velocity map is fairly noisy, we can still discern the effects of disk rotation. The presence of a kinematically distinct C3 component can be interpreted as evidence for disturbed kinematics and/or expanding gas motions. The most redshifted C3 lines are found in position 3, in the centre of region B; here it is separated from the C1 line by 50--80~\kms. C1--C3 velocity differences of $>$50~\kms\ can also be seen in many other regions across the disk.


All three line component maps show again that the axis of rotation of the H$\alpha$-emitting gas is not aligned with the photometric major axis. The offset is such that the most rapidly receding (redshifted) gas is located to the north of the major axis, and the most rapidly approaching (blueshifted) gas lies to the south. As mentioned above, a similar rotation axis tilt was found in the central regions of M82 by \citet[][see their fig.~8a]{r-r04} from observations of the H92$\alpha$ recombination line. We believe that this offset has not previously been noted in the optical, and its detection re-emphasises the advantages of integral field spectroscopy.

\subsubsection{Near-IR emission and absorption line radial velocities} \label{sect:DP_NIR_vel}

\begin{figure*}
\centering
\plotone{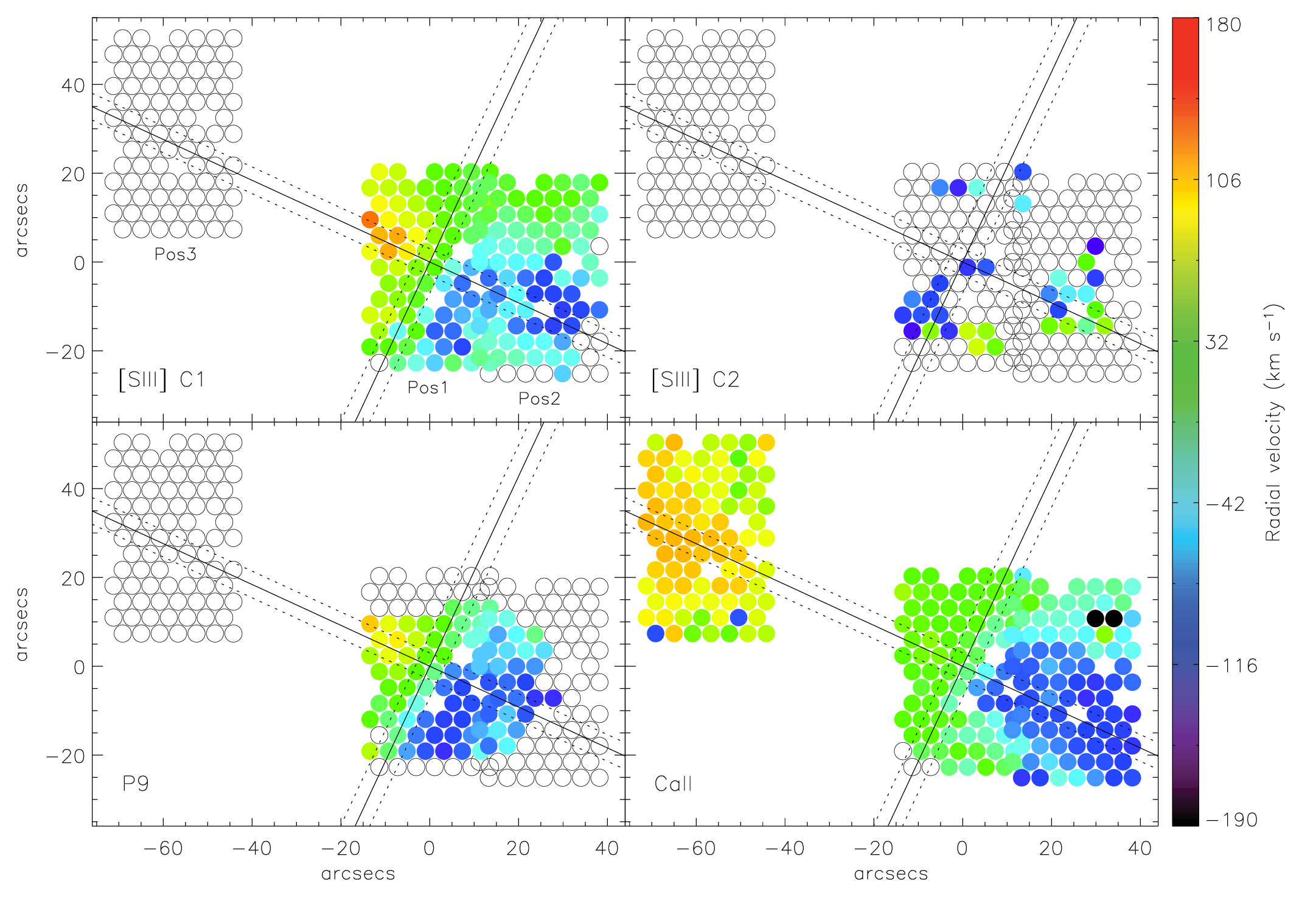} 
\caption{DensePak radial velocity maps in [S\three]$\lambda 9531$ (C1 and C2; upper two panels), P9 (lower-left panel) and Ca\two\ (lower-right panel). The scale is in units of heliocentric \kms\ relative to $v_{\rm sys}$. The galaxy major and minor axes are marked with solid lines, and the dashed lines indicate the width of the pseudo-slits defined in order to make Figs~\ref{fig:dp_major_axis} and \ref{fig:dp_minor_axis}. \textit{[A colour version of this figure is included in the on-line version.]} }
\label{fig:dp_radvel_SIII_CaII}
\end{figure*}

Fig.~\ref{fig:dp_radvel_SIII_CaII} shows radial velocity maps in the [S\three]$\lambda 9531$ emission line (derived from simple Gaussian fits) and the Ca\two\ triplet absorption lines (derived from the pPXF method) of stellar origin.

The [S\three] velocities mirror the H$\alpha$ velocities very well, as expected. Like H$\alpha$, the major axis of rotation appears to be offset with respect to the photometric major axis. In contrast, the Ca\two\ map (lower-right panel) shows that the stellar rotation axis is consistent with the photometric major axis. We measure (by eye) the PA of the gaseous rotation axis to be PA = $53^{\circ}$ \citep[in good agreement with the observations of][]{r-r04}. This is indicated as a dashed line on the C1 map of Fig.~\ref{fig:dp_radvel_Ha}. The offset, therefore, between the photometric/stellar rotation axis and the gaseous rotation axis is $\sim$$12^{\circ}$. Interestingly the molecular CO gas does not exhibit this rotation axis tilt within the central regions, and instead appears to rotate with the same axis as the stars \citep{sofue92, seaquist01, walter02}.

\subsubsection{Major axis position-velocity graph} \label{sect:DP_major}

\begin{figure*}
\centering
\plotone{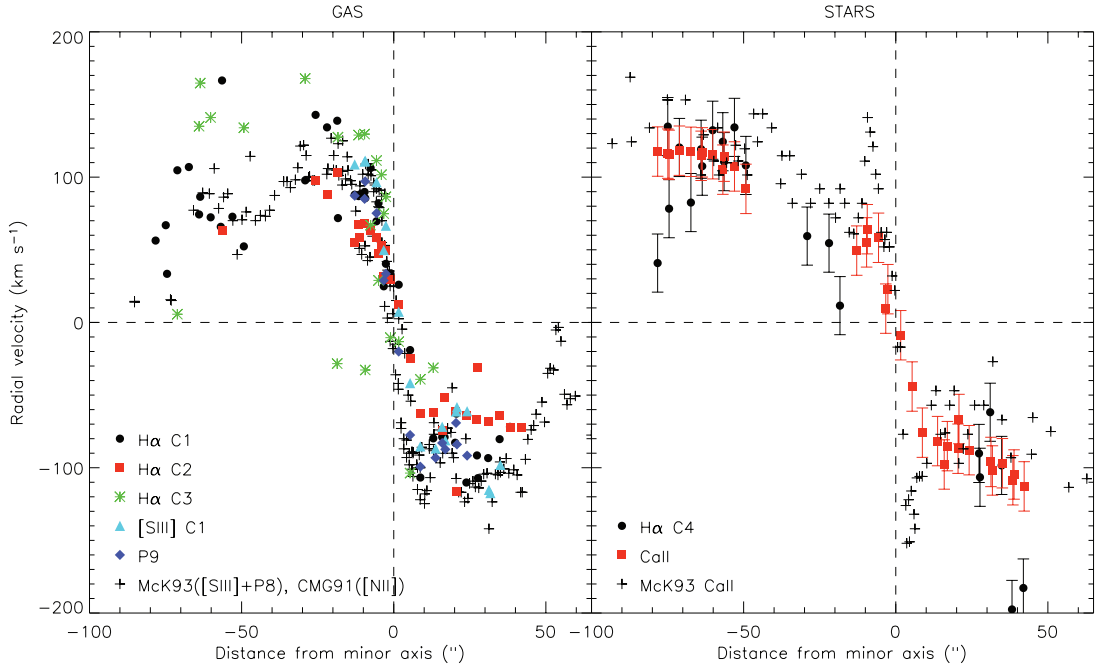} 
\caption{Major-axis position-velocity diagrams for the nebula emission lines (left) and stellar absorption lines (right), extracted from the corresponding $7''$ wide pseudo-slit shown in Figs~\ref{fig:dp_radvel_Ha} and \ref{fig:dp_radvel_SIII_CaII}. Velocities are heliocentrically corrected and shown with respect to $v_{\rm sys}$. On the left panel error bars are not shown, since in most cases they are approximately the same size as the symbols. + symbols represent the optical [N\two]$\lambda$6583 measurements from \citet[][\citetalias{castles91}]{castles91} and the near-IR gas emission line and stellar absorption line measurements from \citetalias{mckeith93}, as indicated. \textit{[A colour version of this figure is included in the on-line version.]} }
\label{fig:dp_major_axis}
\end{figure*}

In order to aid interpretation of these data, we have extracted velocities for each component (emission or absorption) of H$\alpha$, [S\three]$\lambda 9531$, P9 $\lambda 9229$ and Ca\two, within a pseudo-slit $7''$ in width, along the galaxy major-axis. These are shown in Fig.~\ref{fig:dp_major_axis}, where we have split the graph according to the gaseous or stellar origin of the lines. For reference, we have also plotted the [N\two]$\lambda$6583 major axis velocity measurements of \citet[][hereafter \citetalias{castles91}]{castles91} and the [S\three], P8, and Ca\two\ velocities from \citetalias{mckeith93}.

The spatial-resolution of the DensePak IFU is much coarser than previously published long-slit measurements (e.g.\ \citetalias{castles91}, \citetalias{mckeith93}, \citealt{westm07c}) and the GMOS data presented in Section~\ref{sect:GMOS}. We will therefore not discuss these results with respect to the nuclear regions (inner $\pm$$10''$), besides noting that they are in very good agreement. However, the advantage of a much higher spectral resolution and good S/N has meant that we have been able to identify and measure multiple individual line components. We can therefore begin to disentangle some of the complexities of the position-velocity diagram.

A great deal of local variation in the gas velocities is seen, particularly outside the $x_{2}$-orbit  region. For example, at $\sim$$+20''$ there is a dramatic fall in radial velocities, and, with the inclusion of the \citetalias{castles91} and \citetalias{mckeith95} data, we can identify both small- and large-scale ``wiggles'' between $-20''$ and $-80''$. As mentioned above, the shallower velocity gradient of H$\alpha$ C2 (red squares) compared to C1 (black circles) is clearly evident out to $\pm$$50''$. Although C1--C3 line splitting is seen within the central $\pm$$30''$, the most striking result is the 50--80~\kms\ splitting seen at distances of more than $-50''$ from the nucleus (i.e.\ in region B). Clearly this intermediate age region \citep[few 100~Myr;][]{smith07, konstantopoulos08}, which is not part of the current starburst, is still producing high velocity gas.

Our H$\alpha$ stellar absorption velocities agree very well with the Ca\two\ stellar absorption line measurements presented by \citetalias{mckeith93} and \citet{greve02} and the CO velocity measurements of \citet{sofue98}. At the radii at which we can identify the H$\alpha$ absorption component (C4), the corresponding emission-line (H$\alpha$, [S\three], P9) velocities are lower, but again in good agreement with the literature \citepalias{castles91, mckeith93}. The extreme red and blueshifted Ca\two\ velocities at $\pm$$5''$ (associated with the bar $x_{2}$-orbits; \citealt{wills00}; \citetalias{westm07c}) are not observed because our spatial resolution is too coarse. At radii $>$$+10''$, outside of the central $x_{2}$-orbit region, the radial velocities of Ca\two\ exhibit a consistent, but gradual increase (in the blueshifted sense). The equivalent increase on the other (eastern) side of the galaxy is not covered by our DensePak pointings. At radii greater than $-50''$ (for which, again, we do not have equivalent spatial coverage on the opposite side), the Ca\two\ velocities remain very consistent at $\sim$110~\kms. Evidence from \citet{greve02} suggests that beyond this radius, the Ca\two\ velocities remain constant or increase slightly.

The significant differences in stellar and gas radial velocities at radii $<$$-30''$ and $>$$+50''$, together with the fact that the gas rotation curve appears to fall back to zero velocity at +60/-75 arcsecs, can now be understood as resulting from the offset between the gas rotation axis and the photometric and stellar rotation axes. The peak redshifted and blueshifted velocities in the emission line and stellar velocities are numerically very similar, but since they are spatially offset, our pseudo-slit, oriented according to the photometric axis PA, misses the gaseous radial velocity peaks.

\subsubsection{Minor axis position-velocity graph} \label{sect:DP_minor}

\begin{figure*}
\centering
\plotone{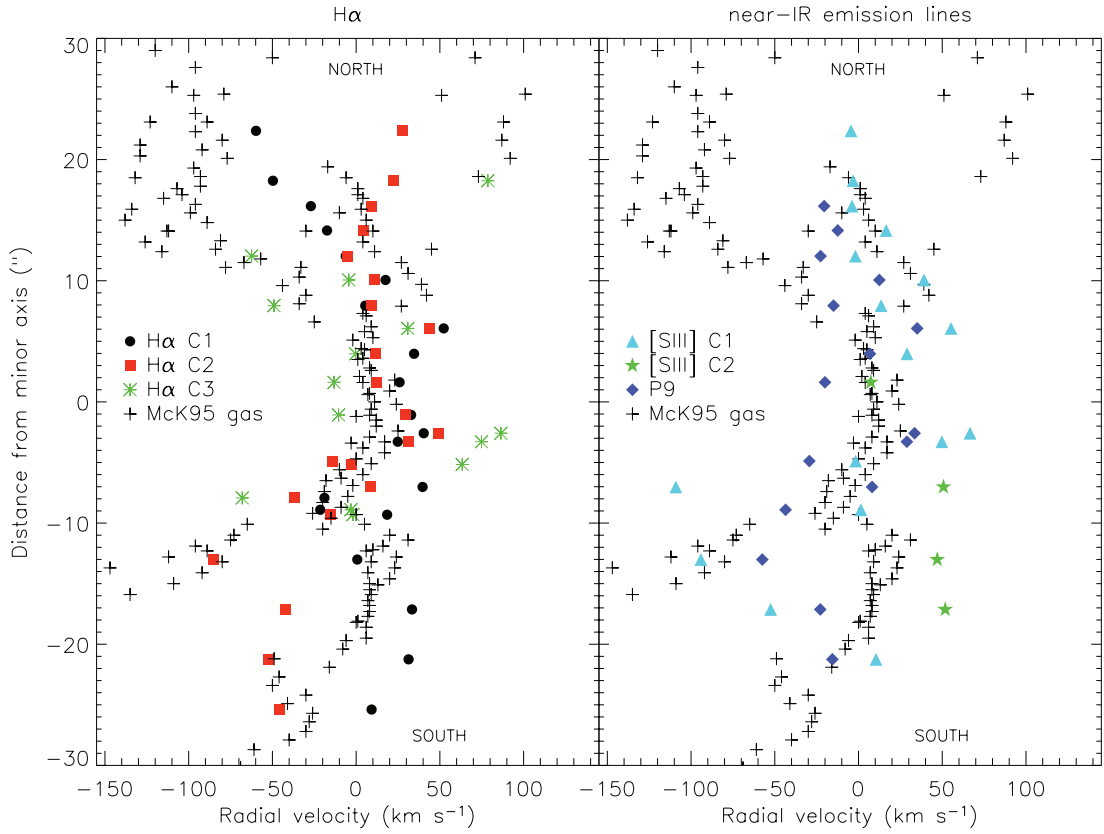} 
\caption{Minor axis position-velocity diagrams for H$\alpha$ (left panel) and [S\three]$\lambda$9531 and P9 (right panel), extracted from the corresponding $7''$ wide pseudo-slit shown in Figs~\ref{fig:dp_radvel_Ha} and \ref{fig:dp_radvel_SIII_CaII}. Velocities are heliocentrically corrected and shown with respect to $v_{\rm sys}$. Error bars have been omitted for clarity; in most cases they are between 5--20~\kms\ with the larger errors applying to the fainter line components. The optical and near-IR gas emission line measurements from \citetalias{mckeith95} are plotted with + symbols. \textit{[A colour version of this figure is included in the on-line version.]} }
\label{fig:dp_minor_axis}
\end{figure*}

Fig.~\ref{fig:dp_minor_axis} shows the minor axis position-velocity graph for the emission line velocities only (split into the H$\alpha$ and near-IR measurements), together with the H$\alpha$, [N\two]$\lambda$6583 and [S\three]$\lambda$9532 data from \citetalias{mckeith95}. The inflection points in the minor-axis position-velocity plot found by \citetalias{mckeith95} lie at $\sim$$\pm$$20''$ ($\pm$350~pc). Recall that, in their interpretation, the region within these markers is the ``energy injection zone'' where the outflow is accelerated to its terminal velocity.

Although our measurements are broadly in agreement with those of \citetalias{mckeith95}, there are a number of differences. For example, in the north we do not see the 100--150~\kms\ blueshifted branch or the 50--100~\kms\ redshifted branch (although there are hints of both in C3). Here C1 of the H$\alpha$ line does become increasingly blueshifted, but only up to $\sim$$60$~\kms\ (at $+23''$). The H$\alpha$ C2 velocities diverge consistently from C1 from $\sim$$12''$, becoming increasingly redder. [S\three] and P9 remain between 0 and $-50$~\kms\ out to $>$$+20''$. Velocities of H$\alpha$ C1 and C2 in the central disk region (within $\pm$$5''$) are fairly consistent with one another, but are, on average, redder than those of \citetalias{mckeith95} by $\sim$20~\kms. Here we identify H$\alpha$ line splitting (C1--C3) of 40--50~\kms. In the south H$\alpha$ C1 broadly follows the red branch of the \citetalias{mckeith95} data, together with [S\three] C2. In this region, [S\three] C1 is redshifted by 100--150~\kms, and appears to be more consistent with the blue branch. Throughout the entire southern region, H$\alpha$ C2 is consistently blueshifted with respect to C1. The differences between our results and that of \citetalias{mckeith95} may arise from a number of reasons, including slight pointing or seeing differences, or our more detailed line profile fitting, including the identification of the broad underlying component, C2, which reveals a more complex situation than could be seen in previous data.

\subsubsection{Summary: DensePak radial velocities}
To summarise this section, firstly we confirm that the axis of gaseous rotation is offset from stellar rotation axis and photometric major axis, and extend this finding to the entire inner 2~kpc of the disk. We measure the offset to be $\sim$12$^{\circ}$. Furthermore, we confirm that the major axis velocity gradient of the H$\alpha$ broad component, C2, is much shallower than that of C1, and that this is true out to $\pm$$50''$ ($\pm$900~pc). 

Line splitting is observed in the narrow component of H$\alpha$ and [S\three] of 40--50~\kms\ throughout the central disk region. Strikingly, we also find splitting in H$\alpha$ of 50--80~\kms\ in the centre of region B. Our velocities along the minor axis are in broad agreement with those found by \citetalias{mckeith95}. In the northern outflow, H$\alpha$ C2 is redshifted with respect to C1, whereas in the south it is blueshifted. The C1--C2 differences range up to 80--90~\kms.

\section{Discussion} \label{sect:disc}


\subsection{The optical emission line components}
As set out in the introduction, we have recently begun to build up a physical model that attempts to explain the origins of the different optical emission line components in winds driven by starbursts based on our observations of the NGC 1569 starburst \citep[][see also \citetalias{westm07c}]{westm07a, westm07b, westm08}. The data presented here for M82 fits well with our ideas, and is of sufficient quality to allow us to develop this basic model further.

As a reminder to the reader, our model accounts for the narrow components of H$\alpha$ (and the other low-ionization optical emission lines), referred to as C1 and C3 (see Paper~II), as originating from the photoionized gas in the disk. This gas is fairly turbulent (typical FWHM 50--120~\kms), and shows copious evidence for localised expanding structures (shells/bubbles) through its split lines (i.e.\ where C3 is observed) and the morphology seen in HST images. In contrast the broad component (C2) represents emission from highly turbulent mixing layers on the surface of denser gas clouds, set up by the impact of high-energy photons and fast-flowing winds from the massive star clusters. Since material is easily evaporated and/or ablated from these turbulent layers, C2 should trace locations of mass-loading sites within the wind.

M82 is a dynamically complex system, perturbed significantly by its gravitational encounter with M81 some $2\times 10^{8}$~yrs ago \citep{yun99}. This complexity, however, is not easily disentangled due to the galaxy's high inclination, strong and variable obscuration, the presence of a bar, and a powerful starburst driving a large-scale outflow. Each sight line towards the centre of the galaxy will traverse considerable material out of the galaxy's midplane before intersecting and possibly passing through a disk/bar star-forming region. Furthermore, both radio recombination line studies and our optical measurements show that the ionized gas and stellar rotation axes are offset by $\sim$12$^{\circ}$. The impact of these geometrical factors, in combination with substantial local structure (e.g. the pseudo-slit `b' in Fig.~\ref{fig:GMOS_maj}), naturally complicate the interpretation of the velocities.

\subsubsection{The narrow components (C1 and C3)}

The generally good agreement between stellar velocities and H$\alpha$ C1 along pseudo-slit `a' (Fig.~\ref{fig:GMOS_maj}) for clump A, implies that C1 originates in the disk of M82. This is also consistent with the presence of compact, bright emission structures in direct narrow band images of clump A that are suggestive of photoionized surfaces in or near the young star clusters (see Fig.~\ref{fig:GMOSfinder}). Additionally, the fact that the widths of H$\alpha$ C1 and C2 remain fairly constant throughout the cores of the starburst clumps is consistent with the known uniformity of the gas excitation and ionization parameter in these regions \citep[][Paper~II]{forster03}, and with the idea of a highly fragmented and well-mixed ISM capable of rapidly adjusting to environmental changes (\citealt{lord96, forster03}; \citetalias{westm07c}).

The detection of C3 indicates strong line-splitting in the narrow component arising from dynamically expanding gaseous structures. Neither finding line splitting throughout the starburst core of M82, nor the comparatively low implied expansion velocities, are unsurprising since this is a very dense but dynamically active region at the roots of the large-scale superwind. In addition since many of these features have small sizes (see Section~\ref{sect:outflow_channel}) they will often crowd along sight lines giving rise to the chaotic appearance and over-turbulent velocities of the M82 disk.   

Our [S\three] line measurements present additional constraints on the emission line models. The differences in dynamics between this higher excitation, near-IR line and the optical emission lines (H$\alpha$, [S\two], etc.) indicate that [S\three] is probing a hotter (and therefore more turbulent) gas phase, co-located with H$\alpha$ C1. This suggests that we may be seeing three distinct gas phases: one originating from the relatively quiescent ionized gas in or near the star clusters (H$\alpha$ -- and [S\two], [N\two], etc.\ -- C1 and C3); one originating from the same place but from a slightly hotter gas phase ([S\three] C1); and one from the turbulent mixing layers on the surfaces of ISM clumps located in moderate-to-low density regions exterior to the starburst core in the inner wind (H$\alpha$ C2).

\subsubsection{The broad component (C2)}

One of the most intriguing results of this study is the difference in disk-wide dynamics found between the narrow and broad ionized gas components. Seen in both our high-resolution GMOS observations, and throughout the inner disk with our DensePak observations, H$\alpha$ C2, in general, exhibits a shallower major axis velocity gradient than C1. Investigating why this is the case is one key to understanding where C2 is emitted from and hence what it represents physically.

Along each sight-line through the galaxy there is a combination of emission due to reflection off dust, as indicated by polarisation measurements \citep[e.g.][]{notni84, scarrott91, jones00}, as well as sources that are seen directly. \citet{bland88} discovered a weak, uniform, broad (300~\kms) emission component in the M82 halo which they associated with scattered light from the nuclear region. Our high resolution, spatially resolved observations indicate that the broad C2 emission may in fact be composed of a local and a scattered light component. However, we believe that the scattered light component must be a minor contributor within the regions we have observed (the starburst core and the inner disk plane) for the following reasons: the C2 velocities vary locally and are imprinted with the large-scale galaxy rotation pattern, and the line widths, densities and line ratios vary locally and with radius (Sections~\ref{sect:GMOS} and \ref{sect:DP}; Paper~II).

A further clue to help us understand the relationship between C1 and C2 comes in the form of the filament described in Section~\ref{sect:GMOS_vel}.

\subsection {Observations of an individual outflow channel} \label{sect:outflow_channel}
Position 5 of the GMOS observations covers a particularly interesting region of the inner outflow -- what appears to be a distinct outflow channel directed away from the starburst clump. By examining the gas properties here, we can begin to better understand the relationship between the (H$\alpha$) narrow and broad components. Along the length of this filament, the narrow component is split with a velocity difference of up to 60~\kms. Lateral cuts across the filament reveal velocity ellipses traced by C1 and C3, whereas C2 appears at rest with respect to the centre of C1-C3 expansion. The edges of the filament are defined by broader C2 widths, higher C2 densities, and low C1 densities ($\sim$700~\cmt\ and $\sim$200~\cmt, respectively; Paper~II), whereas in the centre of the filament, C1 exhibits a higher density ($\sim$700~\cmt, particularly at the end nearest the clumps; Paper~II) and the width of C2 decreases. These facts suggest that this filament is a distinct outflow channel through which hot gas is escaping, interacting with the cooler ISM, and entraining it into the flow. This structure is direct evidence for an interaction between the hot and cool gas phases.

This filament bears a remarkable similarity to to those predicted by \citet{t-t03}, who attempted to model the outflow resulting from multiple, closely spaced super star clusters (SSCs). They found that the interaction of winds from the neighbouring SSCs leads to a network of filaments that originate near the base of the outflow. These structures represent the condensed gas formed through a combination of stationary oblique shocks and crossing shocks that occur between the neighbouring outflows, and may suffer strong radiative cooling depending on the local values of density, temperature and metallicity (and thus emit strongly at both optical and X-ray wavelengths). 

The scale of the M82 starburst and the sheer number of SSCs that it contains raises the question ``why do we not see clear evidence of outflow channels everywhere?''. The answer may lie in the fact that \citeauthor{t-t03}\ model was set up with favourable conditions for the production of coherent channels (the SSCs all had same age and mass, and were distributed at set separations on single plane). The reality of irregular spaced clusters of differing masses and ages is likely to produce a much more chaotic and diffuse flow, although channels similar to the ones predicted clearly have formed.

The high C1 densities near the base of the channel imply that this is where collimation occurs, and the expansion of the narrow component-emitting ``walls'' results in the decrease in this density with radius. The velocities suggest that the C2-emitting gas fills the space between the two walls, and the narrower width of C2 seen in the centre of the channel implies that the interaction between the hot and cool gas is weaker here. For a temperature of $T=10^7$~K (characteristic of the gas associated with the hot wind fluid) the sound speed is $\sim$500~\kms, meaning that the expansion of the channel walls is strongly subsonic (even after taking into account projection effects). This is not unexpected since the surrounding medium is known to be at high pressure \citepalias[][Paper~II]{westm07c}. The identification of this outflow channel therefore provides us with a rare opportunity to gain a deeper insight into the local physical conditions which lead to superwinds, and will be discussed more fully in Paper~II. 
\subsection{The relationship between the narrow and broad line components}
We can now examine in more detail what the relationship between the narrow and broad line components might be, and why C2 appears to rotate at a slower velocity than C1.

If C2 represents turbulent gas stirred up by the interaction between the hot, fast-flowing wind and cooler gas, we would expect C2 to become more dominant as the gas becomes increasingly perturbed and entrained into the outward flow. Thus, if this process is occurring over the length of a channel such as we describe above, then the emission of C1 will be weighted towards the inner-most region (the base of the outflow channel), whereas the C2 emission will be weighted towards regions further out along the flow (i.e.\ extra-planar regions). The velocity gradient difference can therefore be explained, since when observing a highly inclined galaxy, gas located further out in the disk would exhibit a shallower major axis line-of-sight velocity gradient because here the radial (line-of-sight) component of the velocity vector is smaller. The precise situation, however, is complicated by the presence of the bar in M82, since a bar can give rise to a variety of major axis rotation curves (including solid-body-like, or steeper) depending on the bar orientation with respect to the major axis PA \citep{dicaire08}. The outflow channel model also helps explain why the C1 radial velocity maps of the inner wind regions show a large number of localised features that appear to be mirrored by C2 (e.g.\ see pseudo-slit `b' in Fig.~\ref{fig:GMOS_maj}) since the two components are naturally dynamically linked. As the tube expands with radius and continues to rotate with the galaxy disk, the outflow speed of the gas will decrease due to conservation of angular momentum \citep{seaquist01, walter02, greve04}. If the outflow speed is slow enough then the channel will become distorted by these effects; if it is fast enough then the effects will not be significant and the channel will remain straight. Since the filament in our observations appears coherent and linear, we can infer that the outflow channels must be reasonably short lived and contain gas with flow velocities that are large compared to the local rotation speed.


%
\subsection{Implications for the superwind system}
Our GMOS observations sample the gas conditions at the base of the superwind in unprecedented detail. Here the wind is clearly dominated by processes occurring on small spatial scales; these structures are associated with the compressed, fragmentary ISM and the SSCs. Within this inner region, the flow is not collimated as is seen in the outer wind, but fairly dynamically chaotic. Here, fragments of the ISM represent the remnants of interstellar clouds as well as shells or bubbles, shredded by multiple interactions and shocks. Clearly, however, some of this gas is ordered into coherent outflow channels, presumably formed through cluster wind interactions \citep{t-t03} as described above.

Finding C2 emission associated with turbulent clump surface layers and outflow channels throughout the disk of M82 provides powerful direct evidence for the existence of mass-loading, not just at the wind base, but over a large, spatially extended area extending far into the inner wind. The need for large-scale mass-loading in the wind of M82 has been inferred ever since the first analytic wind models \citep{cc85} were compared to X-ray measurements \citep[see also][]{suchkov94}.

The near coexistence of multiple phases of the ISM should lead to complex cooling processes. For example, the measured expansion velocities of the ionized gas features are much less than the the sound speed in the hot gas, implying that the mixed material at the base of the wind is strongly subsonic. These will range from moderate cooling as indicated to fit the bulk X-ray intensity of the wind \citep[e.g.][]{suchkov94} to cases where the dense ISM might locally completely dominate and cool the hot phase.  We also suspect that some fraction of the hot ISM could escape unscathed, and processes associated with this hot, low density component will be difficult to detect in emission.


It has been known for many years that as the ISM gas, stripped and entrained into the wind outflow, reaches greater and greater heights, it retains the signature of disk rotation and exhibits a concomitant decrease in rotation speed. This slowing of rotation above the plane was first observed in CO and then in the optical, and can be (mainly) accounted for by angular momentum conservation \citep{sofue92, seaquist01, walter02, greve04}. Our observations confirm this: the imprint of rotation in the velocity patterns of C1 at all heights is immediately visible in Fig.~\ref{fig:GMOS_min} (through the change in sign of velocity relative to systemic between pseudo-slits 2 and 3 on the east side of M82 versus 4 and 5 on the west) and Fig.~\ref{fig:dp_radvel_Ha}. In agreement with previous studies, these data also suggest that the motions of ISM compressed in expanding bubbles along with entrained matter can help to account for the large mass of extraplanar molecular gas found in M82 \citep[e.g.][]{seaquist01, walter02}.

\section{Summary}\label{sect:summary}

In this paper we have presented two sets of optical integral field spectra of the quintessential starburst galaxy, M82. The spatial coverage, depth and resolution of the combined datasets have given us an unprecedented opportunity for examining this important galaxy. This paper is the first of a series examining the optical structure of M82's disk and its superwind; here we have focussed on the ionized gaseous and stellar dynamics. However, the complexity of the dynamics is not easily disentangled due to the galaxy's high inclination, strong and variable obscuration leading to a degree of scattered emission, the bar and starburst-driven outflow. Our main findings and conclusions are as follows:

\begin{list}{\labelitemi}{\leftmargin=1em}
  \item[$\bullet$] The spectral resolution and S/N of our data have allowed us to examine in detail the shape of the emission line profiles observed. In general, they comprise a bright, narrow component (C1; FWHM $\sim$ 30--130~\kms) superimposed on a broad, fainter component (C2; FWHM $\sim$ 150--350~\kms), with often little or no velocity offset between the two. In some regions two narrow components are identified where the fainter of the two is referred to as C3. Sometimes both narrow components are superimposed on the C2 broad emission component, While C1 is observed across the entirety of our spatial coverage, C2 is identified out to a (plane of the sky) radial extent of 1.7~kpc from the nucleus, and we do not detect it in most of region B.
  \item[$\bullet$] Within the starburst clumps A and C, both the C1 and C2 line widths remain very uniform, and we find localised $\sim$20--70~\kms\ line splitting in the H$\alpha$ narrow component, which we associate with locally expanding shells of compressed, cooler gas that is being photoionized.
  \item[$\bullet$] Narrow C1 and C3 component line splitting of 40--80~\kms\ continues to larger projected distances. Particularly striking is the 50--80~\kms\ splitting seen in region B considering this region is of intermediate age and not part of the current starburst \citep{smith07, konstantopoulos08}. Possibly the few young stars and star clusters in this region suffice to produce shell structures, although the possibility of other mechanisms should not be excluded.
  \item[$\bullet$] Moving out into the inner wind, we find that the line splitting increases to $>$100~\kms\ and both the line splitting and FWHM morphologies begin to follow minor axis oriented patterns. This suggests that here some of the gas (particularly the C2-emitting material) is associated with linear streaming filaments or shocked interfaces. This picture receives further support from the discovery of a distinctive outflow channel described in Section~\ref{sect:outflow_channel}.
  \item[$\bullet$] On average, the narrowest C2 lines are found near the disk midplane. The widths of the C2 line then increase with height above and below the disk to the edges of our spatial coverage. This trend is seen in both H$\alpha$ C2 and [S\three] C1, and is an indication that emission in the broad component becomes more important with increasing height within the energy injection zone.
  \item[$\bullet$] The imprint of disk rotation is seen at all heights meaning that orbital angular momentum is being carried out into the wind \citep[cf.][]{greve04}. The shallower rotation amplitude observed for C2, however, is consistent with it arising at larger heights where conservation of angular momentum has resulted in reduced rotation speeds. This pattern has been previously seen in molecular and ionized gas emission lines observed in the radio where galaxy transparency is not a major issue, and interpreted as axisymmetric drift. Increased importance of the broad C2 optical emission component is also observed moving outward along the outflow channel. The general nature of this phenomenon shows that turbulent gas mixing into outflowing hot gas becomes increasingly important below the main disk of M82.
  \item[$\bullet$] The turn-over in the rotation curve observed on both sides of the galaxy and large C1--C3 line splitting in various places implies that, on average, our line-of-sight extends to at least half-way through the disk, in conflict with the high levels of obscuration usually associated with the nuclear regions of M82.
  \item[$\bullet$] The rotation axis of the gas (as traced by H$\alpha$ C1 and C2 and [S\three] C1) is offset from the stellar rotation axis (as traced by Ca\two\ and H$\alpha$ C4 -- the absorption component) and the photometric major axis PA by $\sim$12$^{\circ}$, not only within the nuclear regions but over the whole inner 2~kpc of the disk. This provides further evidence of its disturbed nature resulting from its interaction within the M81 group.
  \item[$\bullet$] Now that we can identify and map each individual line component, it is clear that this central region, at the base of the large-scale outflow cone, represents a chaotic, complex kinematic environment with many overlapping expanding structures located at different radii, and that the locations at which the narrow component of H$\alpha$ begins to split coherently are found to vary with distance along the major axis. Only at larger radii does the bulk-scale outflow kinematics become more coherent.
  \item[$\bullet$] Our [S\three] observations suggest that we are are probing three distinct gas phases: one originating from the relatively quiescent ionized gas in or near the star clusters (H$\alpha$ C1 and C3); one originating from the same place but from a slightly hotter gas phase ([S\three] C1); and one from the turbulent mixing layers on the surfaces of ISM clumps located in moderate-to-low density regions exterior to the starburst core in the inner wind (H$\alpha$ C2).
\end{list}

The next paper in the series will focus on the nebular characteristics of the ionized gas in the disk and inner wind, including a study of the line ratios, excitations and gas densities. At larger radii, we know from deep H$\alpha$ imaging and spectroscopy that a structured, cone-like outflow develops (\citetalias{mckeith95, shopbell98}; \citealt{ohyama02}). Developing an overall morphological picture of the M82 wind, linking the inner and outer wind regions together, will be the subject of Paper~III.

\section*{Acknowledgments}
We thank the referee for his/her thorough reading of the manuscript and providing many insightful and constructive comments which have led to an improvement in this paper. MSW wishes to thank Kambiz Fathi for interesting discussions relating to galaxy dynamics. JSG's research was partially funded by the National Science Foundation through grant AST-0708967 to the University of Wisconsin-Madison.

Based on observations obtained at the Gemini Observatory, which is operated by the Association of Universities for Research in Astronomy, Inc., under a cooperative agreement with the NSF on behalf of the Gemini partnership: the National Science Foundation (United States), the Science and Technology Facilities Council (United Kingdom), the National Research Council (Canada), CONICYT (Chile), the Australian Research Council (Australia), Minist\'{e}rio da Ci\^{e}ncia e Tecnologia (Brazil) and SECYT (Argentina).

\bibliographystyle{apj}
\bibliography{/Users/msw/Documents/work/references}

\begin{thebibliography}{71}
\expandafter\ifx\csname natexlab\endcsname\relax\def\natexlab#1{#1}\fi

\bibitem[{{Achtermann} \& {Lacy}(1995)}]{achtermann95}
{Achtermann}, J.~M. \& {Lacy}, J.~H. 1995, \apj, 439, 163

\bibitem[{{Allington-Smith} {et~al.}(2002){Allington-Smith}, {Murray},
  {Content}, {Dodsworth}, {Davies}, {Miller}, {Jorgensen}, {Hook},
  {et~al.}}]{allington02}
{Allington-Smith}, J., {Murray}, G., {Content}, R., {Dodsworth}, G., {Davies},
  R., {Miller}, B.~W., {Jorgensen}, I., {Hook}, I., {et~al.} 2002, \pasp, 114,
  892

\bibitem[{{Athanassoula}(1992)}]{athanassoula92a}
{Athanassoula}, E. 1992, \mnras, 259, 328

\bibitem[{{Barden} {et~al.}(1998){Barden}, {Sawyer}, \& {Honeycutt}}]{barden98}
{Barden}, S.~C., {Sawyer}, D.~G., \& {Honeycutt}, R.~K. 1998, in Proc. SPIE
  Vol. 3355, p. 892-899, Optical Astronomical Instrumentation,, ed.
  S.~{D'Odorico}, 892--899

\bibitem[{{Bastian} {et~al.}(2007){Bastian}, {Konstantopoulos}, {Smith},
  {Trancho}, {Westmoquette}, \& {Gallagher}}]{bastian07}
{Bastian}, N., {Konstantopoulos}, I., {Smith}, L.~J., {Trancho}, G.,
  {Westmoquette}, M.~S., \& {Gallagher}, J.~S. 2007, \mnras, 379, 1333

\bibitem[{{Binette} {et~al.}(1999){Binette}, {Cabrit}, {Raga}, \&
  {Cant{\'o}}}]{binette99}
{Binette}, L., {Cabrit}, S., {Raga}, A., \& {Cant{\'o}}, J. 1999, \aap, 346,
  260

\bibitem[{{Bland} \& {Tully}(1988)}]{bland88}
{Bland}, J. \& {Tully}, B. 1988, \nat, 334, 43

\bibitem[{{Burbidge} {et~al.}(1964){Burbidge}, {Burbidge}, \&
  {Rubin}}]{burbidge64}
{Burbidge}, E.~M., {Burbidge}, G.~R., \& {Rubin}, V.~C. 1964, \apj, 140, 942

\bibitem[{{Cappellari} \& {Emsellem}(2004)}]{cappellari04}
{Cappellari}, M. \& {Emsellem}, E. 2004, \pasp, 116, 138

\bibitem[{{Castles} {et~al.}(1991){Castles}, {McKeith}, \& {Greve}}]{castles91}
{Castles}, J., {McKeith}, C.~D., \& {Greve}, A. 1991, Vistas in Astronomy, 34,
  187 (CMG91)

\bibitem[{{Cecil} {et~al.}(2001){Cecil}, {Bland-Hawthorn}, {Veilleux}, \&
  {Filippenko}}]{cecil01}
{Cecil}, G., {Bland-Hawthorn}, J., {Veilleux}, S., \& {Filippenko}, A.~V. 2001,
  \apj, 555, 338

\bibitem[{{Cenarro} {et~al.}(2001){Cenarro}, {Cardiel}, {Gorgas}, {Peletier},
  {Vazdekis}, \& {Prada}}]{cenarro01}
{Cenarro}, A.~J., {Cardiel}, N., {Gorgas}, J., {Peletier}, R.~F., {Vazdekis},
  A., \& {Prada}, F. 2001, \mnras, 326, 959

\bibitem[{{Chevalier} \& {Clegg}(1985)}]{cc85}
{Chevalier}, R.~A. \& {Clegg}, A.~W. 1985, \nat, 317, 44

\bibitem[{{de Vaucouleurs} {et~al.}(1991){de Vaucouleurs}, {de Vaucouleurs},
  {Corwin}, {Buta}, {Paturel}, \& {Fouque}}]{vaucouleurs91}
{de Vaucouleurs}, G., {de Vaucouleurs}, A., {Corwin}, Jr., H.~G., {Buta},
  R.~J., {Paturel}, G., \& {Fouque}, P. 1991, {Third Reference Catalogue of
  Bright Galaxies} (Volume 1-3, XII, 2069~Springer-Verlag Berlin Heidelberg New
  York)

\bibitem[{{Dicaire} {et~al.}(2008){Dicaire}, {Carignan}, {Amram}, {Hernandez},
  {Chemin}, {Daigle}, {de Denus-Baillargeon}, {Balkowski}, {Boselli}, {Fathi},
  \& {Kennicutt}}]{dicaire08}
{Dicaire}, I., {Carignan}, C., {Amram}, P., {Hernandez}, O., {Chemin}, L.,
  {Daigle}, O., {de Denus-Baillargeon}, M.-M., {Balkowski}, C., {Boselli}, A.,
  {Fathi}, K., \& {Kennicutt}, R.~C. 2008, \mnras, 385, 553

\bibitem[{{Dimeo}(2005)}]{dimeo}
{Dimeo}, R. 2005, PAN User Guide

\bibitem[{{F{\"o}rster Schreiber} {et~al.}(2003){F{\"o}rster Schreiber},
  {Genzel}, {Lutz}, \& {Sternberg}}]{forster03}
{F{\"o}rster Schreiber}, N.~M., {Genzel}, R., {Lutz}, D., \& {Sternberg}, A.
  2003, \apj, 599, 193

\bibitem[{{Freedman} {et~al.}(1994){Freedman}, {Hughes}, {Madore}, {Mould},
  {Lee}, {Stetson}, {Kennicutt}, {Turner}, {Ferrarese}, {Ford}, {Graham},
  {Hill}, {Hoessel}, {Huchra}, \& {Illingworth}}]{freedman94}
{Freedman}, W.~L., {Hughes}, S.~M., {Madore}, B.~F., {Mould}, J.~R., {Lee},
  M.~G., {Stetson}, P., {Kennicutt}, R.~C., {Turner}, A., {Ferrarese}, L.,
  {Ford}, H., {Graham}, J.~A., {Hill}, R., {Hoessel}, J.~G., {Huchra}, J., \&
  {Illingworth}, G.~D. 1994, \apj, 427, 628

\bibitem[{{G{\"o}tz} {et~al.}(1990){G{\"o}tz}, {Downes}, {Greve}, \&
  {McKeith}}]{gotz90}
{G{\"o}tz}, M., {Downes}, D., {Greve}, A., \& {McKeith}, C.~D. 1990, \aap, 240,
  52

\bibitem[{{Greve}(2004)}]{greve04}
{Greve}, A. 2004, \aap, 416, 67

\bibitem[{{Greve} {et~al.}(2002){Greve}, {Tarchi}, {H{\"u}ttemeister}, {de
  Grijs}, {van der Hulst}, {Garrington}, \& {Neininger}}]{greve02}
{Greve}, A., {Tarchi}, A., {H{\"u}ttemeister}, S., {de Grijs}, R., {van der
  Hulst}, J.~M., {Garrington}, S.~T., \& {Neininger}, N. 2002, \aap, 381, 825

\bibitem[{{Homeier} \& {Gallagher}(1999)}]{homeier99}
{Homeier}, N.~L. \& {Gallagher}, J.~S. 1999, \apj, 522, 199

\bibitem[{{Izotov} {et~al.}(1996){Izotov}, {Dyak}, {Chaffee}, {Foltz},
  {Kniazev}, \& {Lipovetsky}}]{izotov96}
{Izotov}, Y.~I., {Dyak}, A.~B., {Chaffee}, F.~H., {Foltz}, C.~B., {Kniazev},
  A.~Y., \& {Lipovetsky}, V.~A. 1996, \apj, 458, 524

\bibitem[{{Jones}(2000)}]{jones00}
{Jones}, T.~J. 2000, \aj, 120, 2920

\bibitem[{{Konstantopoulos} {et~al.}(2008){Konstantopoulos}, {Bastian},
  {Smith}, {Trancho}, {Westmoquette}, \& {Gallagher}}]{konstantopoulos08}
{Konstantopoulos}, I.~S., {Bastian}, N., {Smith}, L.~J., {Trancho}, G.,
  {Westmoquette}, M.~S., \& {Gallagher}, III, J.~S. 2008, \apj, 674, 846

\bibitem[{{Larkin} {et~al.}(1994){Larkin}, {Graham}, {Matthews}, {Soifer},
  {Beckwith}, {Herbst}, \& {Quillen}}]{larkin94}
{Larkin}, J.~E., {Graham}, J.~R., {Matthews}, K., {Soifer}, B.~T., {Beckwith},
  S., {Herbst}, T.~M., \& {Quillen}, A.~C. 1994, \apj, 420, 159

\bibitem[{{Lester} {et~al.}(1990){Lester}, {Gaffney}, {Carr}, \&
  {Joy}}]{lester90}
{Lester}, D.~F., {Gaffney}, N., {Carr}, J.~S., \& {Joy}, M. 1990, \apj, 352,
  544

\bibitem[{{Lord} {et~al.}(1996){Lord}, {Hollenbach}, {Haas}, {Rubin}, {Colgan},
  \& {Erickson}}]{lord96}
{Lord}, S.~D., {Hollenbach}, D.~J., {Haas}, M.~R., {Rubin}, R.~H., {Colgan},
  S.~W.~J., \& {Erickson}, E.~F. 1996, \apj, 465, 703

\bibitem[{{Lynds} \& {Sandage}(1963)}]{lynds63}
{Lynds}, C.~R. \& {Sandage}, A.~R. 1963, \apj, 137, 1005

\bibitem[{{Marlowe} {et~al.}(1995){Marlowe}, {Heckman}, {Wyse}, \&
  {Schommer}}]{marlowe95}
{Marlowe}, A.~T., {Heckman}, T.~M., {Wyse}, R.~F.~G., \& {Schommer}, R. 1995,
  \apj, 438, 563

\bibitem[{{McKeith} {et~al.}(1993){McKeith}, {Castles}, {Greve}, \&
  {Downes}}]{mckeith93}
{McKeith}, C.~D., {Castles}, J., {Greve}, A., \& {Downes}, D. 1993, \aap, 272,
  98 (McK93)

\bibitem[{{McKeith} {et~al.}(1995){McKeith}, {Greve}, {Downes}, \&
  {Prada}}]{mckeith95}
{McKeith}, C.~D., {Greve}, A., {Downes}, D., \& {Prada}, F. 1995, \aap, 293,
  703 (McK95)

\bibitem[{{Melo} {et~al.}(2005){Melo}, {Mu{\~n}oz-Tu{\~n}{\'o}n},
  {Ma{\'{\i}}z-Apell{\'a}niz}, \& {Tenorio-Tagle}}]{melo05}
{Melo}, V.~P., {Mu{\~n}oz-Tu{\~n}{\'o}n}, C., {Ma{\'{\i}}z-Apell{\'a}niz}, J.,
  \& {Tenorio-Tagle}, G. 2005, \apj, 619, 270

\bibitem[{{Mendez} \& {Esteban}(1997)}]{mendez97}
{Mendez}, D.~I. \& {Esteban}, C. 1997, \apj, 488, 652

\bibitem[{{Notni} \& {Bronkalla}(1984)}]{notni84}
{Notni}, P. \& {Bronkalla}, W. 1984, Astronomische Nachrichten, 305, 157

\bibitem[{{O'Connell} {et~al.}(1995){O'Connell}, {Gallagher}, {Hunter}, \&
  {Colley}}]{oconnell95}
{O'Connell}, R.~W., {Gallagher}, III, J.~S., {Hunter}, D.~A., \& {Colley},
  W.~N. 1995, \apj, 446, L1

\bibitem[{{O'Connell} \& {Mangano}(1978)}]{oconnell78}
{O'Connell}, R.~W. \& {Mangano}, J.~J. 1978, \apj, 221, 62

\bibitem[{{Ohyama} {et~al.}(2002){Ohyama}, {Taniguchi}, {Iye}, {Yoshida},
  {Sekiguchi}, {Takata}, {Saito}, {Kawabata}, {et~al.}}]{ohyama02}
{Ohyama}, Y., {Taniguchi}, Y., {Iye}, M., {Yoshida}, M., {Sekiguchi}, K.,
  {Takata}, T., {Saito}, Y., {Kawabata}, K.~S., {et~al.} 2002, \pasj, 54, 891

\bibitem[{{Pettini} {et~al.}(2001){Pettini}, {Shapley}, {Steidel}, {Cuby},
  {Dickinson}, {Moorwood}, {Adelberger}, \& {Giavalisco}}]{pettini01}
{Pettini}, M., {Shapley}, A.~E., {Steidel}, C.~C., {Cuby}, J.-G., {Dickinson},
  M., {Moorwood}, A.~F.~M., {Adelberger}, K.~L., \& {Giavalisco}, M. 2001,
  \apj, 554, 981

\bibitem[{{Pittard} {et~al.}(2005){Pittard}, {Dyson}, {Falle}, \&
  {Hartquist}}]{pittard05}
{Pittard}, J.~M., {Dyson}, J.~E., {Falle}, S.~A.~E.~G., \& {Hartquist}, T.~W.
  2005, \mnras, 361, 1077

\bibitem[{{Rodriguez-Rico} {et~al.}(2004){Rodriguez-Rico}, {Viallefond},
  {Zhao}, {Goss}, \& {Anantharamaiah}}]{r-r04}
{Rodriguez-Rico}, C.~A., {Viallefond}, F., {Zhao}, J.-H., {Goss}, W.~M., \&
  {Anantharamaiah}, K.~R. 2004, \apj, 616, 783

\bibitem[{{Sait\={o}} {et~al.}(1984){Sait\={o}}, {Sasaki}, {Kaneko},
  {Nishimura}, \& {Toyama}}]{saito84}
{Sait\={o}}, M., {Sasaki}, M., {Kaneko}, N., {Nishimura}, M., \& {Toyama}, K.
  1984, \pasj, 36, 305

\bibitem[{{Sawyer}(1997)}]{sawyer97}
{Sawyer}, D. 1997, DensePak Users Manual

\bibitem[{{Scarrott} {et~al.}(1991){Scarrott}, {Eaton}, \& {Axon}}]{scarrott91}
{Scarrott}, S.~M., {Eaton}, N., \& {Axon}, D.~J. 1991, \mnras, 252, 12P

\bibitem[{{Schmidt} {et~al.}(1976){Schmidt}, {Angel}, \&
  {Cromwell}}]{schmidt76}
{Schmidt}, G.~D., {Angel}, J.~R.~P., \& {Cromwell}, R.~H. 1976, \apj, 206, 888

\bibitem[{{Seaquist} \& {Clark}(2001)}]{seaquist01}
{Seaquist}, E.~R. \& {Clark}, J. 2001, \apj, 552, 133

\bibitem[{{Shapley} {et~al.}(2003){Shapley}, {Steidel}, {Pettini}, \&
  {Adelberger}}]{shapley03}
{Shapley}, A.~E., {Steidel}, C.~C., {Pettini}, M., \& {Adelberger}, K.~L. 2003,
  \apj, 588, 65

\bibitem[{{Shen} \& {Lo}(1995)}]{shen95}
{Shen}, J. \& {Lo}, K.~Y. 1995, \apj, 445, L99

\bibitem[{{Shopbell} \& {Bland-Hawthorn}(1998)}]{shopbell98}
{Shopbell}, P.~L. \& {Bland-Hawthorn}, J. 1998, \apj, 493, 129 (SBH98)

\bibitem[{{Sidoli} {et~al.}(2006){Sidoli}, {Smith}, \& {Crowther}}]{sidoli06}
{Sidoli}, F., {Smith}, L.~J., \& {Crowther}, P.~A. 2006, \mnras, 370, 799

\bibitem[{{Slavin} {et~al.}(1993){Slavin}, {Shull}, \& {Begelman}}]{slavin93}
{Slavin}, J.~D., {Shull}, J.~M., \& {Begelman}, M.~C. 1993, \apj, 407, 83

\bibitem[{{Smith} {et~al.}(2007){Smith}, {Bastian}, {Konstantopoulos},
  {Gallagher}, {Gieles}, {de Grijs}, {Larsen}, {O'Connell}, \&
  {Westmoquette}}]{smith07}
{Smith}, L.~J., {Bastian}, N., {Konstantopoulos}, I.~S., {Gallagher}, III,
  J.~S., {Gieles}, M., {de Grijs}, R., {Larsen}, S.~S., {O'Connell}, R.~W., \&
  {Westmoquette}, M.~S. 2007, \apjl, 667, L145

\bibitem[{{Smith} {et~al.}(2006){Smith}, {Westmoquette}, {Gallagher},
  {O'Connell}, {Rosario}, \& {de Grijs}}]{smith06}
{Smith}, L.~J., {Westmoquette}, M.~S., {Gallagher}, J.~S., {O'Connell}, R.~W.,
  {Rosario}, D.~J., \& {de Grijs}, R. 2006, \mnras, 370, 513

\bibitem[{{Sofue} {et~al.}(1992){Sofue}, {Reuter}, {Krause}, {Wielebinski}, \&
  {Nakai}}]{sofue92}
{Sofue}, Y., {Reuter}, H.-P., {Krause}, M., {Wielebinski}, R., \& {Nakai}, N.
  1992, \apj, 395, 126

\bibitem[{{Sofue} {et~al.}(1998){Sofue}, {Tomita}, {Tutui}, {Honma}, \&
  {Takeda}}]{sofue98}
{Sofue}, Y., {Tomita}, A., {Tutui}, Y., {Honma}, M., \& {Takeda}, Y. 1998,
  \pasj, 50, 427

\bibitem[{{Stevens} {et~al.}(2003){Stevens}, {Read}, \&
  {Bravo-Guerrero}}]{stevens03a}
{Stevens}, I.~R., {Read}, A.~M., \& {Bravo-Guerrero}, J. 2003, \mnras, 343, L47

\bibitem[{{Strickland} \& {Heckman}(2007)}]{strickland07}
{Strickland}, D.~K. \& {Heckman}, T.~M. 2007, \apj, 658, 258

\bibitem[{{Suchkov} {et~al.}(1994){Suchkov}, {Balsara}, {Heckman}, \&
  {Leitherer}}]{suchkov94}
{Suchkov}, A.~A., {Balsara}, D.~S., {Heckman}, T.~M., \& {Leitherer}, C. 1994,
  \apj, 430, 511

\bibitem[{{Telesco} {et~al.}(1991){Telesco}, {Joy}, {Dietz}, {Decher}, \&
  {Campins}}]{telesco91}
{Telesco}, C.~M., {Joy}, M., {Dietz}, K., {Decher}, R., \& {Campins}, H. 1991,
  \apj, 369, 135

\bibitem[{{Tenorio-Tagle} {et~al.}(2003){Tenorio-Tagle}, {Silich}, \&
  {Mu{\~n}oz-Tu{\~n}{\'o}n}}]{t-t03}
{Tenorio-Tagle}, G., {Silich}, S., \& {Mu{\~n}oz-Tu{\~n}{\'o}n}, C. 2003, \apj,
  597, 279

\bibitem[{{van Dokkum}(2001)}]{vandokkum01}
{van Dokkum}, P.~G. 2001, \pasp, 113, 1420

\bibitem[{{Vanzi} {et~al.}(2006){Vanzi}, {Scatarzi}, {Maiolino}, \&
  {Sterzik}}]{vanzi06}
{Vanzi}, L., {Scatarzi}, A., {Maiolino}, R., \& {Sterzik}, M. 2006, \aap, 459,
  769

\bibitem[{{Walter} {et~al.}(2002){Walter}, {Wei{\ss}}, \&
  {Scoville}}]{walter02}
{Walter}, F., {Wei{\ss}}, A., \& {Scoville}, N. 2002, \apjl, 580, L21

\bibitem[{{Wei{\ss}} {et~al.}(2001){Wei{\ss}}, {Neininger}, {H{\"u}ttemeister},
  \& {Klein}}]{weis01}
{Wei{\ss}}, A., {Neininger}, N., {H{\"u}ttemeister}, S., \& {Klein}, U. 2001,
  \aap, 365, 571

\bibitem[{{Westmoquette} {et~al.}(2007{\natexlab{a}}){Westmoquette}, {Exter},
  {Smith}, \& {Gallagher}}]{westm07a}
{Westmoquette}, M.~S., {Exter}, K.~M., {Smith}, L.~J., \& {Gallagher}, J.~S.
  2007{\natexlab{a}}, \mnras, 381, 894

\bibitem[{{Westmoquette} {et~al.}(2008){Westmoquette}, {Smith}, \&
  {Gallagher}}]{westm08}
{Westmoquette}, M.~S., {Smith}, L.~J., \& {Gallagher}, J.~S. 2008, \mnras, 383,
  864

\bibitem[{{Westmoquette} {et~al.}(2007{\natexlab{b}}){Westmoquette}, {Smith},
  {Gallagher}, \& {Exter}}]{westm07b}
{Westmoquette}, M.~S., {Smith}, L.~J., {Gallagher}, J.~S., \& {Exter}, K.~M.
  2007{\natexlab{b}}, \mnras, 381, 913

\bibitem[{{Westmoquette} {et~al.}(2007{\natexlab{c}}){Westmoquette}, {Smith},
  {Gallagher}, {O'Connell}, {Rosario}, \& {de Grijs}}]{westm07c}
{Westmoquette}, M.~S., {Smith}, L.~J., {Gallagher}, III, J.~S., {O'Connell},
  R.~W., {Rosario}, D.~J., \& {de Grijs}, R. 2007{\natexlab{c}}, \apj, 671, 358 (W07c)

\bibitem[{{Wills} {et~al.}(2000){Wills}, {Das}, {Pedlar}, {Muxlow}, \&
  {Robinson}}]{wills00}
{Wills}, K.~A., {Das}, M., {Pedlar}, A., {Muxlow}, T.~W.~B., \& {Robinson},
  T.~G. 2000, \mnras, 316, 33

\bibitem[{{Yun}(1999)}]{yun99}
{Yun}, M.~S. 1999, in IAU Symposium, Vol. 186, Galaxy Interactions at Low and
  High Redshift, ed. J.~E. {Barnes} \& D.~B. {Sanders}, 81

\end{thebibliography}



\clearpage

\clearpage

\clearpage

\clearpage

\clearpage

\clearpage

\clearpage

\clearpage

\clearpage

\clearpage

\clearpage

\clearpage

\clearpage

\clearpage

\clearpage

\clearpage

\clearpage

\clearpage

\clearpage

\clearpage

\clearpage

\end{document}